\documentclass[12pt,a4paper, table, xcdraw]{article}
\pdfoutput=1
\usepackage{geometry}
\UseRawInputEncoding
\usepackage{amsmath}
\usepackage{amsmath}
\usepackage{amssymb}
\usepackage[dvips]{graphicx}
\usepackage{cite}
\usepackage{pstricks}
\usepackage{bm}
\usepackage{bbm}
\usepackage{pbox}
\usepackage{placeins}
\usepackage{graphicx}
\usepackage{caption}
\usepackage{subcaption}
\usepackage[T1]{fontenc}
\usepackage{footnote}
\usepackage{pdfpages}
\usepackage{hhline}
\usepackage{multirow}
\usepackage{multicol}
\usepackage{enumitem}
\usepackage{multirow}
\usepackage{amsmath}
\usepackage{relsize}
 \usepackage{graphicx}
\usepackage[toc,page]{appendix}
\usepackage[english]{babel}
\allowdisplaybreaks
\usepackage{xcolor}
\usepackage{float}
\usepackage{slashed}
\usepackage{comment}
\usepackage{cases}
\usepackage{empheq}
\usepackage[normalem]{ulem}

\definecolor{MyDarkBlue}{rgb}{0.1, 0.3, 0.8} 
\definecolor{SBlue}{rgb}{0.2, 0.4, 0.4} 
\definecolor{MyLightBlue}{rgb}{0.22,0.51,0.99}
\definecolor{MyGreen}{rgb}{0.0, 0.5, 0.3}
\definecolor{BrickRed}{rgb}{0.8, 0.25, 0.33}
\usepackage[colorlinks=true,linkcolor=blue,citecolor=MyDarkBlue,
urlcolor=BrickRed,bookmarksnumbered=true,bookmarksopen]{hyperref}
\hypersetup{colorlinks, citecolor=SBlue,linkcolor=MyGreen, urlcolor=BrickRed}

\begin{document}
\vspace*{-0.2in}
\begin{flushright}
\end{flushright}
\begin{center}
{\Large \bf
Exploiting a future galactic supernova to probe\\\vspace{0.05 in}  neutrino magnetic moments
}
\end{center}
\renewcommand{\thefootnote}{\fnsymbol{footnote}}
\begin{center}
{
{}~\textbf{Sudip Jana,$^1$}\footnote{ E-mail: \textcolor{MyDarkBlue}{sudip.jana@mpi-hd.mpg.de}}
{}~\textbf{Yago P Porto-Silva,$^2$}\footnote{ E-mail: \textcolor{MyDarkBlue}{yporto@ifi.unicamp.br}}
{}~\textbf{Manibrata Sen$^1$}\footnote{ E-mail: \textcolor{MyDarkBlue}{manibrata@mpi-hd.mpg.de}}
}
\vspace{0.3cm}
{
\\\em$^1$Max-Planck-Institut f{\"u}r Kernphysik, Saupfercheckweg 1, 69117 Heidelberg, Germany
\\
$^2$Instituto de F{\'i}sica Gleb Wataghin - UNICAMP, 13083-859, \\ Campinas, S\~ao Paulo, Brazil
} 
\end{center}
\renewcommand{\thefootnote}{\arabic{footnote}}
\setcounter{footnote}{0}
\thispagestyle{empty}

\begin{abstract}
A core-collapse supernova (SN) offers an excellent astrophysical laboratory to test non-zero neutrino magnetic moments. In particular, the neutronization burst phase, which lasts for few tens of milliseconds post-bounce, is dominated by electron neutrinos and can offer exceptional discovery potential for transition magnetic moments. We simulate the neutrino spectra from the burst phase in forthcoming neutrino experiments like the Deep Underground Neutrino Experiment (DUNE), and the Hyper-Kamiokande (HK), by taking into account spin-flavour conversions of supernova neutrinos caused by interactions with ambient magnetic fields. We find that the neutrino transition magnetic moments which can be explored by these experiments for a galactic SN are an order to several orders of magnitude better than the current terrestrial and astrophysical limits. Additionally, we also discuss how this realization might provide light on three important neutrino properties: (a) the Dirac/Majorana nature, (b) the neutrino mass ordering, and (c) the neutrino mass-generation mechanism.

\end{abstract}
\newpage
\setcounter{footnote}{0}

{
  \hypersetup{linkcolor=black}
  \tableofcontents
}
\newpage

\section{Introduction}\label{SEC-01}
The discovery of neutrino oscillations, driven by non-zero neutrino masses and mixing, provided the first robust evidence of physics beyond the Standard Model (SM)~\cite{ParticleDataGroup:2020ssz}. In order to account for the tiny neutrino masses and mixing, the SM must be extended. In these extensions, a generic outcome of mechanisms of neutrino mass generation is the existence of a non-zero neutrino magnetic moment $(\mu_{\nu})$ through quantum loop corrections. Thereby, a careful theoretical and experimental study of neutrino electromagnetic interactions may help shed some light on the underlying theory. Within the SM minimally extended to contain Dirac neutrino masses, neutrino magnetic moments are $\mu_{\nu}\lesssim 10^{-19}$ $\mu_B$ ~\cite{PhysRevLett.45.963}, where $\mu_B=e/(2 m_e)$ is the Bohr magneton, $e$ is the electron charge, and $m_e$ is the electron mass. On the other hand, the strength of transition neutrino magnetic moments, in case of Majorana scenario in the standard seesaw mechanism, is even smaller and of the order of $\sim 10^{-23}~\mu_B$~\cite{Pal:1981rm}. 
Such a small values $\mu_\nu$ is far beyond the reach of current experimental capabilities.

New symmetries and interactions \cite{Voloshin:1987qy, Babu:1989wn, Babu:2020ivd, Barr:1990um, Georgi:1990za} can change the above picture, and allow for larger values of $\mu_\nu$, without getting into unrealistic values of neutrino masses (for a review, see Ref.~\cite{Giunti:2014ixa}). Neutrino magnetic moments can either be flavour-diagonal and/or flavour off-diagonal, depending on the nature of the neutrino. Dirac neutrinos can have an intrinsic magnetic moment that gives rise to spin precession for a given flavour $\alpha$, i.e. $\nu_{\alpha L} \rightarrow \nu_{\alpha R}$ (where $L,R$ refer to left-handed helicity and right-handed helicity respectively) in the presence of sufficiently strong magnetic fields~\cite{PhysRevLett.45.963}, as well as a transition magnetic moment (TMM) that induces flavour precession in addition to rotation of the spin, i.e., $\nu_{\alpha L} \rightarrow \nu_{\beta R}$~\cite{Pal:1981rm,Shrock:1982sc,Bilenky:1987ty}. For Dirac neutrinos, the latter processes would cause the resultant neutrinos to be invisible since they will have the wrong helicity to be detected through weak interactions. Within the minimally extended SM, such TMMs of Dirac neutrinos are small due to a cancellation as a result of the Glashow-Iliopolus-Maiani (GIM) mechanism. On the other hand, Majorana neutrinos can have only transition moments, and the resultant spin-flavour precession would convert a neutrino into an antineutrino of a different flavour, and vice-versa~\cite{Shrock:1982sc}. Furthermore, coherent forward scattering of neutrinos off the surrounding matter background can resonantly enhance these spin-flavour conversions, thereby termed as the resonant spin-flavour precession (RSFP)~\cite{Akhmedov:1987nc, Lim:1987tk, Akhmedov:1988uk}. This is very similar to matter enhancement of neutrino oscillations due to the Mikheyev-Smirnov-Wolfenstein (MSW) effect~\cite{PhysRevD.17.2369, Mikheev:1986gs}.

Bounds on neutrino magnetic moments can be placed using neutrinos from a variety of terrestrial as well as astrophysical sources.
The search for the neutrino magnetic moments began seven decades ago \cite{Cowan:1954pq}, even before the neutrino was discovered. However, it became a very popular topic three decades ago when the Chlorine experiment found an apparent time variation of solar neutrino flux in anti-correlation with the Sun-spot activity \cite{Davis:1988gd, Davis:1990fb}. After that, several neutrino experiments (reactor-based experiments like  KRASNOYARSK \cite{Vidyakin:1992nf},  ROVNO  \cite{Derbin:1993wy}, MUNU  \cite{Daraktchieva:2005kn}, TEXONO \cite{Deniz:2009mu}, GEMMA \cite{Beda:2012zz} and CONUS \cite{Bonet:2022imz}; solar neutrino experiment like Borexino \cite{Borexino:2017fbd}; accelerator-based experiments like LAPMF \cite{Allen:1992qe} and  LSND experiment~\cite{LSND:2001akn}) investigated neutrino magnetic moments by looking at $\nu_e-e$ scattering in general. The best constraint from these lab-based experiments, coming from Borexino, sets $\mu_{\nu} < 2.8 \times 10^{-11}~\mu_B$ \cite{Borexino:2017fbd}. The GEMMA \cite{Beda:2012zz} and CONUS \cite{Bonet:2022imz} collaborations placed somewhat less stringent constraints on neutrino magnetic moments of the order of  $\mathcal{O} (10^{-11})~ \mu_B.$ Other terrestrial experiments mentioned above have shown sensitivity to neutrino magnetic moments of the order of  $\mathcal{O} (10^{-10})~ \mu_B$. The topic of neutrino magnetic moments started receiving renewed attention when the XENON1T experiment~\cite{Aprile:2020tmw} observed an excess of low energy electron recoil events, which can be explained by a neutrino transition magnetic moment $\mu_{\nu} \in(1.4,2.9) \times 10^{-11} \mu_{B}$ at $90 \%$ C.L.

Strong bounds on neutrino magnetic moment can arise from astrophysical setups as well~\cite{Akhmedov:1992ea, Athar:1995cx,Nunokawa:1998vh}. The presence of a non-zero neutrino magnetic moment allows for a direct coupling between neutrinos and photons, thereby allowing for neutrino radiative decays, as well as plasmon decays to neutrino-antineutrino pairs. The strongest bounds usually arise from globular cluster stars, where plasmon decay can delay helium ignition, leading to anomalous cooling of stars. Absence of any such observational evidence leads to $\mu_{\nu}\leq 3\times 10^{-12} \mu_B$~\cite{Raffelt:1999tx}. For an updated bound, check~\cite{Capozzi:2020cbu}. Similar cooling bounds also exist from observations of neutrinos from SN1987A. The presence of a non-zero neutrino magnetic moment allows for $\nu_L\rightarrow \nu_R$ conversions, leading to further cooling, and hence a shorter duration of neutrino emission than observed. Constraining the amount of $\nu_R$ produced, one arrives at $\mu_\nu \leq 10^{-11}\, \mu_B$~\cite{PhysRevLett.61.27}.  Cosmological constraints on neutrino magnetic moments arise from observation of the primordial abundances of elements during Big Bang Nucleosynthesis (BBN), which sets constraints of the order of $\mathcal{O} (10^{-10}) ~ \mu_B$~\cite{Vassh:2015yza}.

The neutrino flux from a core-collapse supernova (SN) can be an excellent tool to constrain neutrino magnetic moments. Neutrino flavour evolution inside a SN is currently an unsolved problem in the realm of particle physics. The dense matter environment, coupled with strong magnetic fields, makes a SN ideal for studying flavour transitions due to RSFP~\cite{Lim:1987tk, Akhmedov:1988uk}. This is specially relevant for Majorana neutrinos, where non-zero TMMs can lead to $\nu_e \rightarrow \bar{\nu}_{x}$, where $\nu_x$ is linear combination of the non-electron flavours, $\nu_\mu$ and $\nu_\tau$. Related studies were performed in~\cite{Ando:2002sk, Ando:2003pj, Ando:2003is, Akhmedov:2003fu}, where this mechanism was shown to mix different neutrino flavours, causing spectral changes. A simple, effective two-flavour analysis of neutrino flavour evolution in the presence of non-zero TMMs identifies,  in addition to the usual MSW resonances, two extra resonances: (i) the RSFP-H, associated with the atmospheric mass-squared difference $(\Delta m^2_{31})$, occurring at high densities, and (ii) the RSFP-L, associated with the solar mass-squared difference $(\Delta m^2_{21})$, occurring at lower densities. Ref.~\cite{Akhmedov:2003fu} studied the impact of such RSFP in a three-flavour setup and discussed an additional type of resonance, the RSFP-E (electron-type), which is present in a pure three-flavour analysis. The RSFP-E can result in converting $\nu_e \rightarrow \bar{\nu}_x$ in the normal mass ordering (NO), while in the inverted mass ordering (IO), it would result in $\bar{\nu}_e \rightarrow \nu_x$. In all these cases, it was found that all the resonances associated with spin-flavour precession depend very sensitively on the electron number fraction per nucleon, denoted by $Y_e$. These results can be further complicated due to the presence of collective flavour oscillations, arising out of the non-linearity of the equations of motion associated with neutrino flavour evolution within a SN~\cite{deGouvea:2012hg,deGouvea:2013zp}.

However, flavour transitions due to RSFP will not be very effective if the spectra of different neutrino flavours are similar. This is mostly true of the cooling phase -- a period existing for $\mathcal{O}(1-10)$s post-bounce. However, the first $\mathcal{O}(30)$ms post-bounce is dominated by the deleptonization epoch, also known as the neutronization burst (described in detail in the next section). This epoch is characterized by a pure $\nu_e$ flux, with very little contamination of $\bar{\nu}_e$ and $\nu_{\mu,\tau}$. This phase is a robust feature of all hydrodynamic simulations, and hence acts as an excellent laboratory to test new neutrino properties. As a result, any new physics mechanism that ends up converting $\nu_e$s would cause a dramatic change in the $\nu_e$ spectra. In particular, it was discussed in Ref~\cite{Akhmedov:2003fu,Ando:2003is} that combination of MSW and RSFP transitions can result in $\nu_e \rightarrow \bar{\nu}_e$ conversions during this epoch. A smoking-gun signal of such conversion would be a suppression of the $\nu_e$ flux while leading to an enhancement in the $\bar{\nu}_e$ channel, which can be detected in experiments sensitive to neutrinos from a SN.

In this work, we expand on this idea to utilize the neutronization burst from a future galactic SN to put some of the strongest constraints on neutrino TMMs. We simulate the neutrino spectra from the burst phase in upcoming neutrino experiments such as the Deep Underground Neutrino Experiment (DUNE)~\cite{Acciarri:2016crz,DUNE:2020zfm}, and the Hyper-Kamiokande (HK)~\cite{Abe:2018uyc}. Since DUNE has maximum sensitivity to the $\nu_e$ spectra, while HK can measure the $\overline{\nu}_e$, both these experiments can be used to constrain TMMs of neutrinos using the neutronization burst spectra. The large sizes of these detectors and their detection channels make them ideal for probing the effect of TMM on the burst spectra. We find that the values of neutrino TMM that can be probed by these experiments for a SN occurring at 10$\,$kpc are two or three orders of magnitude better than the current terrestrial as well as astrophysical constraints. We perform a parameterized study of the effect of neutrino TMM in this paper. In order to relate it to the fundamental parameters of a given model, we consider a simple model based on $SU(2)_{ H}$ horizontal symmetry \cite{Babu:2020ivd}, which can give rise to large neutrino magnetic moments. We translate the bounds obtained from our analysis onto the model parameters.

The paper is structured as follows: in the next section, we briefly discuss the supernova neutronization burst. Then we analyze in detail the neutrino flavour conversion in the presence of neutrino magnetic moments. Subsequently, we analyze the discovery potential of neutrino magnetic moments in forthcoming experiments, and finally, we discuss the implications on neutrino properties before we conclude.

\section{Supernova neutronization burst}\label{SEC-03}
The supernova neutronization burst is a period of deleptonization of the SN core, which lasts for about 30~ms after core-bounce~\cite{Janka:2006fh}. During collapse, as the core stiffens and reaches nuclear densities, a shockwave is launched, which travels outwards dissociating the surrounding nuclei on its way into corresponding nucleons. Electron capture on protons leads to a large burst of $\nu_e$, which dominates the spectral content during this period, with subdominant contributions from $\bar{\nu}_e$ and $\nu_{\mu,\tau}$. 
\begin{figure}[!t]
  \centering
  \includegraphics[width=0.5\textwidth]{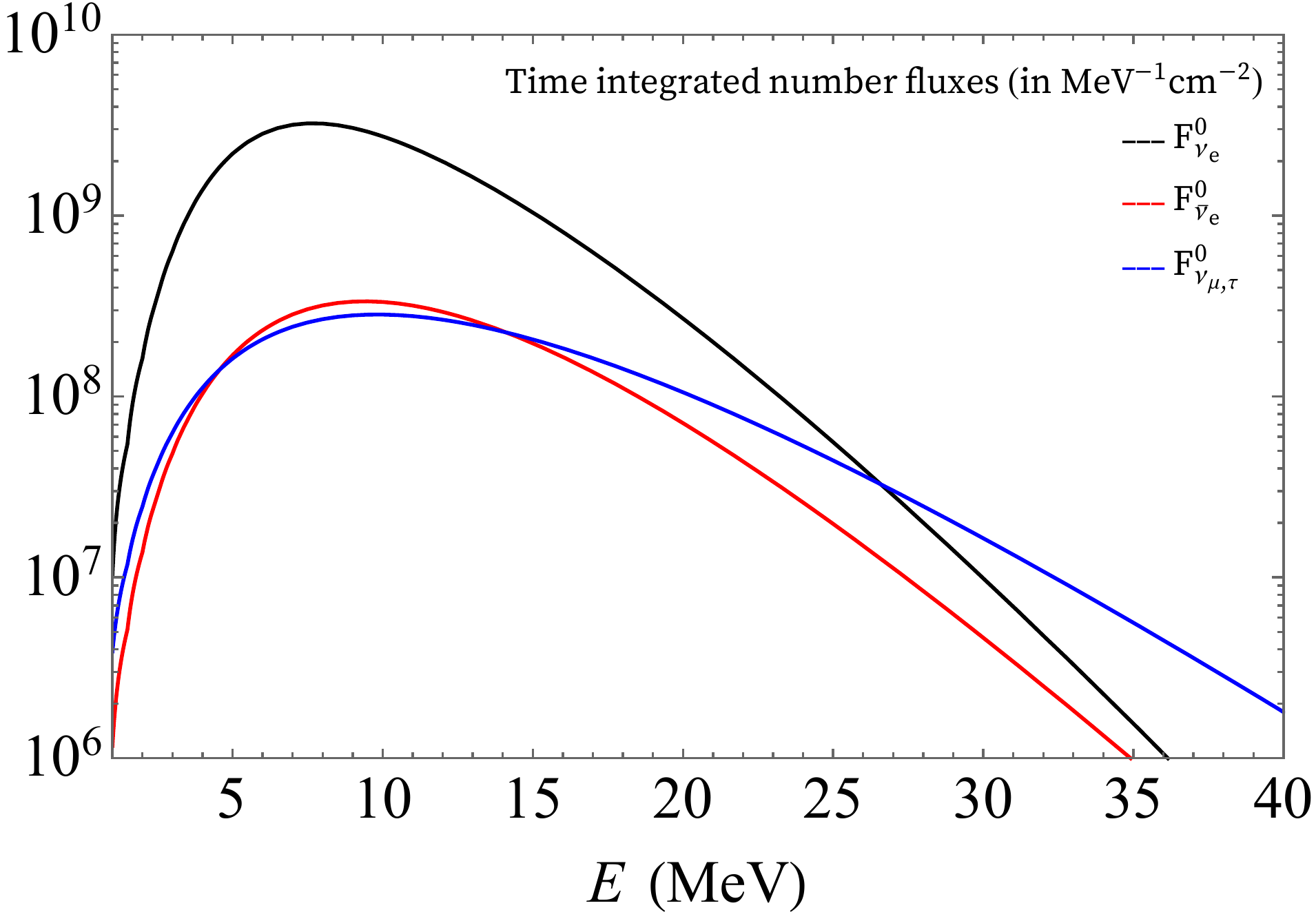}
  \caption{Time-integrated neutrino number fluxes (upto $50\,$ms) for the neutronization burst epoch. Note that $\nu_e$ dominates over other flavours upto $E\lesssim 25\,$MeV.}
  \label{fig:Neut_initial_flux}
\end{figure}
The neutrino spectra from a SN can be parameterized using the ``alpha-fit'' spectra~\cite{Keil:2002in}: 
 \begin{equation}
  F^0_{\nu}(E)=\frac{L_{\nu}}{\langle E_\nu \rangle^2}\frac{(\alpha+1)^{(\alpha+1)}}{\Gamma (\alpha+1)}\left(\frac{E}{\langle E_\nu \rangle}\right)^{\alpha}{\rm exp}\left[-(\alpha+1) \frac{E}{\langle E_\nu \rangle}\right]\,,
  \label{spectra_ch1}
 \end{equation}
where $L_\nu$ is the neutrino luminosity, $\langle E_\nu\rangle$ is the average energy of the neutrino, $\Gamma(z)$ denotes the Euler gamma function, and the pinching parameter, $\alpha$, defined as,
\begin{equation}
 \frac{1}{1+\alpha}=\frac{\langle E^2\rangle-\langle E\rangle^2}{\langle E\rangle^2}\,.
\label{eq:alpha}
\end{equation}
is related to the spectral width. 
Fig.~\ref{fig:Neut_initial_flux} shows the time-integrated (upto $t=50\,$ms) neutrino spectra during this period, obtained from a spherically symmetric simulation of a $15\,M_\odot$ progenitor~\cite{Garching}. Evidently, this period is dominated by the $\nu_e$ spectra for $E\leq 25\,$MeV, while the non-electron flavours dominate above that energy. 

Deep inside the SN, the matter densities are high enough that the $\nu_e$ is produced as the instantaneous Hamiltonian eigenstate associated with the largest eigenvalue. During evolution, the flavour eigenstates decohere, and hence can be treated as an incoherent superposition of mass eigenstates. Assuming adiabatic evolution inside the SN, the $\nu_e$ flux at the Earth can be written as~\cite{Dighe:1999bi}
\begin{equation}
F_{\nu_e}(E) = \frac{1}{4\pi R^2}\left[|{\rm U}_{eh}|^2\, F^0_{\nu_e}(E) +
(1-|{\rm U}_{eh}|^2)\, F^0_{\nu_{\mu,\tau}}(E) \right]
\label{eq:NueEarth}
\end{equation}
where $U_{eh}$ is the corresponding flavour element of the PMNS matrix, and $h$ denotes the heaviest mass eigenstate. In the NO, the $\nu_e$ produced inside the SN comes out of it as a $\nu_3$, whereas in the IO, it comes out as a $\nu_2$. Since the other neutrino flavours are sub-dominant during this epoch, the flux in the NO is suppressed with respect to that in the IO by a factor of $\sim |{\rm U}_{e3}|^2/ |{\rm U}_{e2}|^2 \simeq 0.1 $. In principle, this suppression of the flux in the NO as compared to the IO can be utilized to probe the neutrino mass ordering~\cite{Dighe:1999bi}. For antineutrinos, since the sign of the matter potential inverts, a $\bar{\nu}_e$ produced will exit the SN as the lightest mass eigenstate. However, this entire picture changes as soon as neutrinos acquire a non-trivial magnetic moment, as we will discuss in the next section.
\section{Neutrino flavour Conversion in the presence of magnetic moments}\label{SEC-04}
Neutrinos can have Dirac and/or Majorana magnetic moments, which can be parameterized by the following interaction Lagrangians,
\begin{gather}
    \mathcal{L}_{D} \supset  \mu_{D} \overline{\nu_{L}} \sigma_{\eta \delta} \nu_{R} F^{\eta \delta}\\ 
  \mathcal{L}_{M} \supset  \mu_{\alpha\beta} \nu_{\alpha L}^{T} C \sigma_{\eta \delta} \nu_{\beta L}^{} F^{\eta \delta}
\end{gather}
where $F^{\eta \delta}$ is the electromagnetic field-strength tensor, and $\mu_{D}$ and $\mu_{\alpha\beta}$ denote the Dirac (or intrinsic), and Majorana (or transition) magnetic moment operators. Here $\nu_\alpha$ and $\nu_\beta$ denote neutrino flavours $\alpha$ and $\beta$ respectively, since transition magentic moments (TMM) connect neutrinos of different flavours. In the presence of a magnetic field in the exploding star, such TMM can lead to $\nu_{\alpha} \rightarrow \bar{\nu}_{\beta}$ transition, which violates lepton number by 2 units. On the other hand, the  intrinsic magnetic moment (IMM) can only allow for $\nu_{\alpha L} \rightarrow \nu_{\alpha R}$ to occur, without changing the neutrino flavour. 
For Majorana neutrinos, only the TMMs  $(\alpha\neq \beta)$ are non-zero, while the  IMMs $\mu_{\alpha\alpha}$ are zero\footnote{Due to $CPT$ invariance, if $\alpha = \beta $, the Majorana moment will be exactly zero. A particle's magnetic moment should be equal and opposite to that of its anti-particle, but since a Majorana particle is its own anti-particle, its diagonal moments must disappear. This does not preclude off-diagonal transition magnetic moments.}. 

In this section we focus on TMMs of Majorana neutrinos, and study their flavour evolution as they propagate radially out in the magnetic fields of the exploding star. In such situations, neutrinos undergo several resonances driven by the matter enhanced MSW, as well as spin-flavour precession transitions. The interplay between RSFP and subsequent MSW conversions along the SN profile brings an interesting possibility where the $\nu_e$ flux during the neutronization burst can be efficiently converted into $\bar{\nu}_e$ giving a very distinctive signal in experiments sensitive to SN neutrinos.

\subsection{Evolution equation}
We consider a three-flavour setup, and study the evolution of neutrinos and antineutrinos in the presence of a finite transverse magnetic field intensity $B_{\perp}$. Ignoring temporal variations, the Equation of Motion (EoM) in matrix form is given by,
\begin{equation} \label{evolution}
    i \frac{d}{d r}\left(\begin{array}{c}\nu \\ \bar{\nu}\end{array}\right)=\left(\begin{array}{cc}H_{\nu} & B_{\perp} M \\ -B_{\perp} M & H_{\bar{\nu}}\end{array}\right)\left(\begin{array}{c}\nu \\ \bar{\nu}\end{array}\right),
\end{equation}
where $r$ is the radial coordinate in the SN. The neutrino flavour dynamics is governed by the Hamiltonian,
\begin{equation}
    H_{\nu}=\frac{1}{2 E} U\left(\begin{array}{ccc}0 & 0 & 0 \\ 0 & \Delta m_{21}^{2} & 0 \\ 0 & 0 & \Delta m_{31}^{2}\end{array}\right) U^{\dagger}+\left(\begin{array}{ccc}V_{\nu_e} & 0 & 0 \\ 0 & V_{\nu_\mu} & 0 \\ 0 & 0 & V_{\nu_\tau}\end{array}\right),
\end{equation}
where $U$ is the PMNS matrix, $E$ is neutrino energy and $\Delta m^2_{ij}$ gives the mass square difference between vacuum eigenstates $\nu_i$ and $\nu_j$, and $V$ denotes the forward scattering potential due to coherent scattering of neutrinos on electrons and nucleons. Assuming equal number densities of electron and protons ($n_e=n_p$), the potentials, at leading order, have the following form,
\begin{equation}
    V_{\nu_e}=\sqrt{2}G_F(n_e-\frac{1}{2}n_n) \hspace{0.5cm} \text{and} \hspace{0.5cm} V_{\nu_{\mu,\tau}}=-\frac{1}{2}\sqrt{2}G_Fn_n,
\end{equation}
where $n_e$ and $n_n$ represent the electron and neutron number densities, respectively. The corresponding Hamiltonian for the antineutrinos, $H_{\bar{\nu}}$, only differs in the sign of the potential, such that $V_{\bar{\nu}_e}=-V_{\nu_e}$ and $V_{\bar{\nu}_{\mu,\tau}}=-V_{\nu_{\mu,\tau}}$. Note that the relevant quantity for the flavour evolution is the difference of the $\nu_e$ and $\nu_{\mu,\tau}$ potentials, since an overall phase term can always be rotated away.

The matrix $M$ contains the TMMs:
\begin{equation} \label{M-matrix}
    M=\left(\begin{array}{ccc}0 & \mu_{e \mu} & \mu_{e \tau} \\ -\mu_{e \mu} & 0 & \mu_{\mu \tau} \\ -\mu_{e \tau} & -\mu_{\mu \tau} & 0\end{array}\right).
\end{equation}
The three flavours of neutrinos and anti-neutrinos are grouped together in a column given by
\begin{equation}
    \nu=\left(\begin{array}{c}\nu_{e} \\ \nu_{\mu} \\ \nu_{\tau}\end{array}\right), \quad \bar{\nu}=\left(\begin{array}{c}\bar{\nu}_{e} \\ \bar{\nu}_{\mu} \\ \bar{\nu}_{\tau}\end{array}\right).
\end{equation}
It is  the $B_{\perp} M$ term that couples the neutrinos to the antineutrinos, and gives rise to RSFP conversions. Note that in this analysis, we neglect the neutrino self-interaction terms, which are known to cause collective oscillations among different flavours~\cite{Duan:2006an,Hannestad:2006nj}. This is primarily because the negligible proportions of $\bar{\nu}_e$ and $\nu_{\mu,\tau}$ suppresses collective oscillations during the neutronization epoch.

\begin{figure}[!t]
  \centering
  \includegraphics[width=0.5\textwidth]{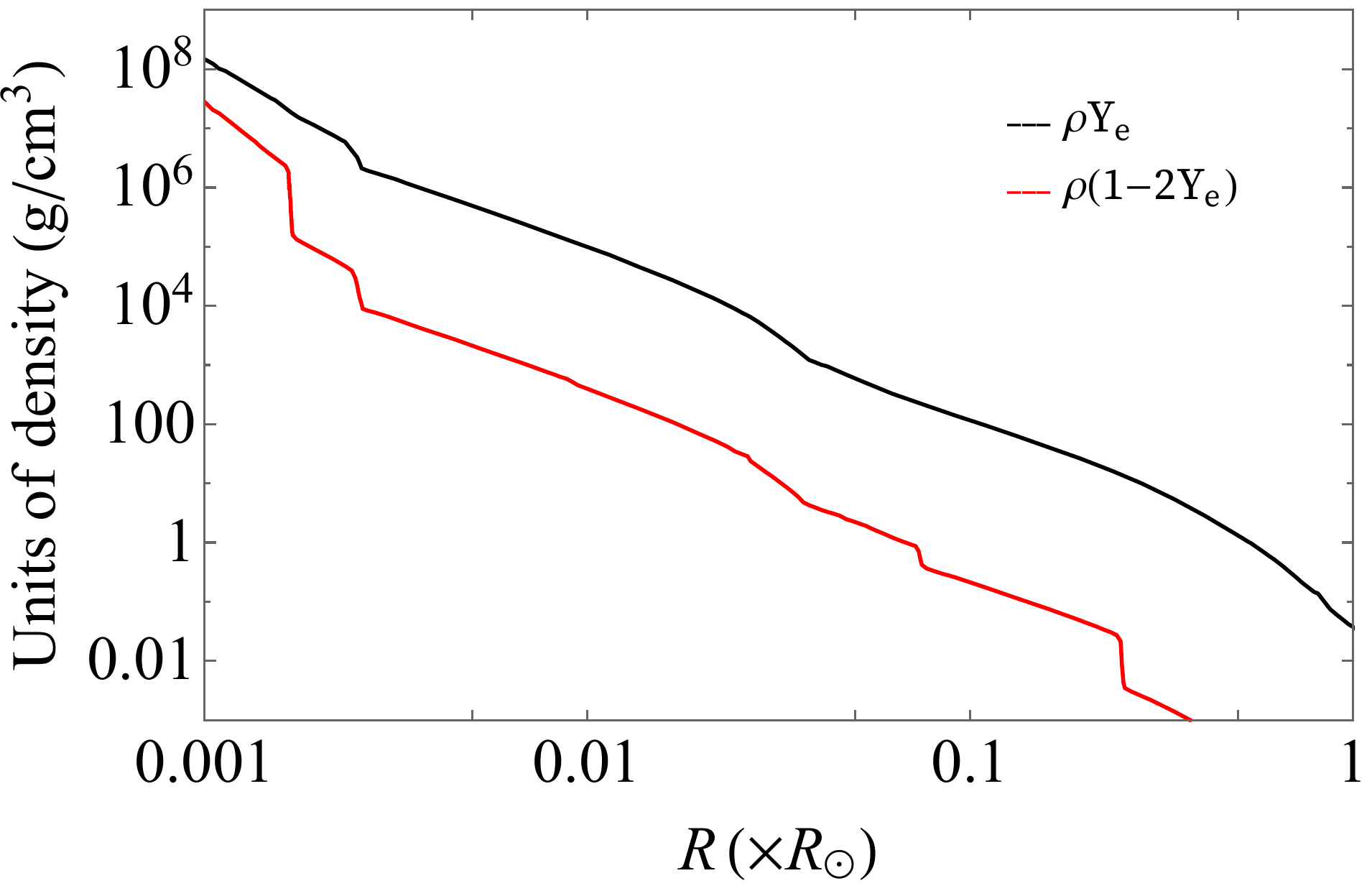}
  \caption{The quantities, $\rho(1-2Y_e)$ and $\rho Y_e$, relevant for RSFP and MSW transitions respectively. These are taken from the static profile of a $15 M_\odot$ star with solar metallicity~\cite{Woosley:1995ip}. Here we consider $R_\odot=696340$ Km.}
   \label{fig:profile}
\end{figure}

We rewrite the EoMs in terms of the electron number fraction per nucleon $Y_e=n_e/(n_p+n_n)$, such that the electron number density $n_e=Y_e \rho/m_N$, where $\rho$ is the matter density and $m_N$ is the nucleon mass. 
For our discussions with regards to resonant flavour conversions, the relevant potential difference for the MSW flavour conversion is
\begin{equation} \label{MSW-potential}
    V_{\nu_e}-V_{\nu_{\mu,\tau}}=\sqrt{2}G_F \frac{\rho}{m_N} Y_e=-(V_{\bar{\nu}_e}-V_{\bar{\nu}_{\mu,\tau}}).
\end{equation}
whereas for the RSFP flavour conversions, it is
\begin{equation} \label{RSFP-potential}
    V_{\nu_e}-V_{\bar{\nu}_{\mu,\tau}}=-\sqrt{2}G_F \frac{\rho}{m_N} (1-2Y_e)=-(V_{\bar{\nu}_e}-V_{\nu_{\mu,\tau}}).
\end{equation}

Clearly, in order to proceed with the neutrino flavour evolution, it is important to know the matter profile $\rho$, as well as the 
electron number fraction, $Y_e$. We use the static profile model for a progenitor mass $15 M_\odot$ and solar metallicity as shown in Fig.\,\ref{fig:profile}~\cite{Woosley:1995ip}. We focus on the almost isotopically neutral region where $(1-2Y_e)$ is small.

The magnetic field within a SN can be parameterized as a dipole,
\begin{equation} \label{mag-field}
    B_{\perp}=B_0\left(\frac{r_0}{r}\right)^3,
\end{equation}
where $r_0$ is the radius of the iron core, which is set to $r_0\approx 0.0024 R_\odot$ for a progenitor mass $15 M_\odot$~\cite{Totani:1996wf}. The magnitude of the magnetic field at the equator of the iron surface is given by $B_0$, and its value can be deduced from the observation of white dwarfs to be in the range $10^7$ G $\lesssim B_0 \lesssim 10^{10}$ G \cite{Totani:1996wf}.

\subsection{Level crossing scheme} \label{level-crossing}
To understand the neutrino flavour evolution, it is useful to work in the basis in which the $\nu_\mu-\nu_\tau$ and $\bar{\nu}_\mu-\bar{\nu}_\tau$ subspaces are simultaneously diagonal. The flavour states in the new basis are primed: $\nu_{\mu}'$, $\nu_{\tau}'$, $\bar{\nu}_{\mu}'$ and $\bar{\nu}_{\tau}'$. In addition, the magnetic moment elements in Eq.\,\ref{M-matrix} alter accordingly to the change of basis and are called $\mu_{e \mu'}$ and $\mu_{e \tau'}$. For simplicity, we set $\mu_{\mu \tau}=0$ as all our results are insensitive to it\footnote{For a detailed discussion of the bounds on $\mu_{\mu \tau}$, see  \cite{Guzzo:2012rf}}.

We compute the energy levels of the instantaneous matter eigenstates of the Hamiltonian in Eq.\,\ref{evolution} for both mass orderings and show them in Fig.\,\ref{fig:LevCros}. The values of the mixing parameters were artificially chosen for illustration purpose in Fig.~\ref{fig:LevCros}, but the arrangements of the resonances is in good agreement with calculation using current mixing best fit values.  For very high densities, $\rho \gtrsim 10^6$ $\text{g/cm}^3$, the matter eigenstates coincide with the flavour states in the primed basis. Both diagrams display five resonances: the usual MSW-H (converts $\nu_{e} \rightleftharpoons \nu_{\tau}^{\prime}$ for NO and $\bar{\nu}_{e} \rightleftharpoons \bar{\nu}_{\tau}^{\prime}$ for IO), and MSW-L (converts $\nu_{e} \rightleftharpoons \nu_{\mu}^{\prime}$ for both orderings), and three additional spin-flavour resonances, RSFP-H ($\bar{\nu}_{e} \rightleftharpoons \nu_{\tau}^{\prime}$ for NO and $\nu_{e} \rightleftharpoons \bar{\nu}_{\tau}^{\prime}$ for IO), RSFP-L ($\bar{\nu}_{e} \rightleftharpoons \nu_{\mu}^{\prime}$ for NO and IO) and RSFP-E ($\nu_{e} \rightleftharpoons \bar{\nu}_{\mu}^{\prime}$ for NO and $\bar{\nu}_{e} \rightleftharpoons \nu_{\mu}^{\prime}$ for IO). Out of these, the MSW-H, MSW-L, RSFP-H, and RSFP-L can be estimated using the two-flavour approximation. The condition for the  MSW-H(L) is given by
\begin{equation}
    \sqrt{2}G_F \frac{\rho}{m_N} Y_e \approx \frac{|\Delta m^2_{j1}|}{2E}\cos \theta_{1j}
\end{equation}
where $j=3$ for H and $j=2$ for L. For RSFP-H(L), the corresponding resonance condition is
\begin{equation}
    \sqrt{2}G_F \frac{\rho}{m_N} (1-2Y_e) \approx \frac{|\Delta m^2_{j1}|}{2E}\cos \theta_{1j}.
\end{equation}

On the other hand, the RSFP-E only appears in a full three-flavour analysis~\cite{Akhmedov:2003fu}. Furthermore, note that the RSFP-H(L) happen at higher densities than MSW-H(L), because the former is driven by $(1-2Y_e)\lesssim 10^{-3}$, whereas the latter is governed only by $Y_e \approx 0.5$, hence the densities need to be higher in RSFP-H(L).

The corresponding adiabaticity parameter for the MSW and RSFP resonances are
\begin{equation} \label{ad-MSW}
    \gamma_{\text{MSW-H(L)}}=\frac{\sin ^{2} 2 \theta_{1 j}}{\cos 2 \theta_{1 j}} \frac{\Delta m_{j 1}^{2}}{2 E} \left|\frac{1}{\rho Y_e} \frac{d\left(\rho Y_e\right)}{d r}\right|^{-1},
\end{equation}
and
\begin{equation} \label{ad-RSFP}
    \gamma_{\text{RSFP-H(L)}} \simeq \frac{8 E}{\Delta m_{j1}^{2}}\left(\mu_{e \beta'} B_{\perp }\right)^{2} \left|\frac{1}{\rho (1-2Y_e)} \frac{d\left(\rho (1-2Y_e)\right)}{d r}\right|^{-1},
\end{equation}
where $\beta=\mu$ for L and $\beta=\tau$ for H.
Eqs.\,(\ref{ad-MSW}) and (\ref{ad-RSFP}) must be evaluated at the resonance point and adiabaticity is guaranteed for $\gamma \gg 1$. When neutrinos encounter an adiabatic resonance, the transitions between different energy levels in Fig.~\ref{evolution} are negligible. If adiabaticity is not met, a instantaneous matter eigenstate follows the dashed lines in Fig.~\ref{evolution} and hops from its energy level to a close one with probability,
\begin{equation} \label{hopping}
    P_{\rm res}\approx e^{-\frac{\pi}{2} \gamma_{\rm res}},
\end{equation}
where $\gamma_{\rm res}$ assumes the form in Eqs.~(\ref{ad-MSW}) or (\ref{ad-RSFP}) with the subscript ${\rm res}$ referring to the specific resonance. In the non-adiabatic limit, $\gamma_{\rm res} \ll 1$, the hopping probability in (\ref{hopping}) approaches unity and the matter eigenstates corresponding the two energy levels involved in the resonance exchange their previous fluxes.
For lack of precise analytical expression for the adiabaticity of RSFP-E \cite{Akhmedov:2003fu}, we rely on numerical numerical calculations.
In what follows, we discuss these resonances in detail for both mass orderings.

\subsubsection{Normal mass ordering (NO)}
The analysis is simplified because of the adiabaticity of the MSW-H and MSW-L resonances. We find, numerically, for 
\begin{equation} \label{muB-small}
    \mu_\nu  B_0  \lesssim 10^{-2} \mu_B \text{G}\,,
\end{equation}
the RSFP-L and RSFP-E are completely non-adiabatic. On the other hand, the RSFP-H can transit from completely non-adiabatic to adiabatic in the interval
\begin{equation} \label{muB-range}
    0.5 \times 10^{-3} \mu_B \text{G} \lesssim \mu_\nu  B_0  \lesssim 10^{-2} \mu_B \text{G}\,.
\end{equation}
The closer $\mu_\nu  B_0$ is to $10^{-2} \mu_B \text{G}$, the more adiabatic is the RSFP-H, whereas for values $\gtrsim 10^{-2}\mu_B \text{G}$, adiabaticity  at RSFP-H saturates to a maximum. The probability that transitions happen between energy levels of neutrinos due to non-adiabaticity of the RSFP-H is given by $p_H$ (i.e., $p_H=P_{\text{RSFP-H}}$, see Eq.~(\ref{hopping})).

\begin{figure} [!t]
  \includegraphics[width=1\textwidth]{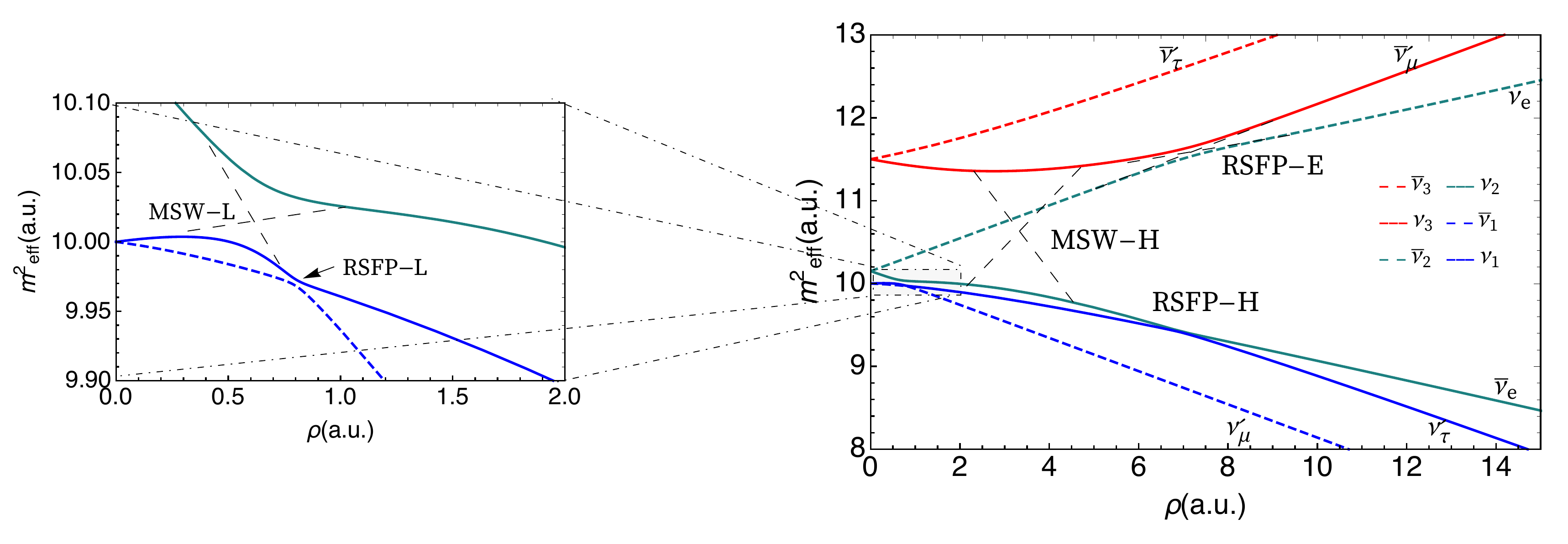}\\
  \vspace{0.1in}
  \includegraphics[width=1\textwidth]{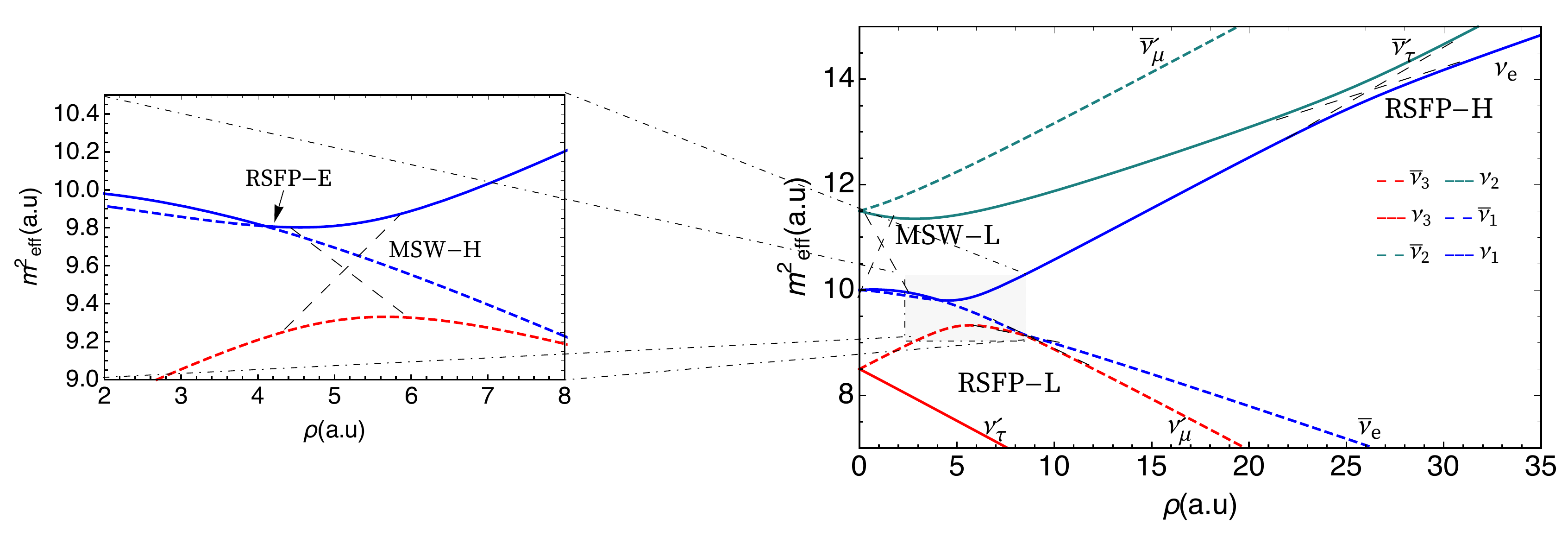}
  \caption{Energy level crossing diagrams for neutrinos and antineutrinos for the Hamiltonian considered in Eq.\,(\ref{evolution}).  The top panel shows the crossings for NO, while the bottom panel shows the same for IO. The axes are arbitrarily scaled to highlight the resonances.}
  \label{fig:LevCros}
\end{figure}
The energy levels of matter eigenstates in case of NO are given in the top panel of Fig.\,\ref{fig:LevCros}. The coloured lines indicate the evolution of the matter eigenstates (solid for neutrinos, dashed for antineutrinos), whereas the dashed black lines track the flavour content. We find that in the presence of non-zero magnetic moment deep inside the star, the $\nu_e$ state coincides with initial $\bar{\nu}_2$, and $\bar{\nu}_\mu'$ with $\nu_3$, therefore $F_{\bar{\nu}_2}=F_{\nu_e}^0$, whereas $F_{\nu_3}=F_{\bar{\nu}_\mu'}^0$. As the non-adiabatic RSFP-E resonance is hit, the flux associated with the matter eigenstates $\nu_2$ and $\nu_3$ switch, thereby $F_{\nu_e}^0 = F_{\nu_3}$ and $F_{\bar{\nu}_\mu'}^0 = F_{\bar{\nu}_2}$.
The next resonance encountered by $\nu_3$ is the MSW-H, which is adiabatic, while $\bar{\nu}_2$ finds no other resonances, and hence the neutrino states exit the SN as
\begin{eqnarray}
   F_{\nu_3}&=&F_{\nu_e}^0,\,\\
   F_{\bar{\nu}_2} &=& F_{\bar{\nu}_\mu'}^0.
\end{eqnarray}
The $\bar{\nu}_e$ is produced deep inside as a $\nu_2$, whereas the $\nu_\tau'$ is produced mainly as a $\nu_1$. As these states evolve, they encounter the RSFP-H, where the $\nu_2$ can flip to the $\nu_1$ and vice versa with a probability $p_H$. At this stage, the $\nu_2$ and $\nu_1$ fluxes can be written as an admixture of the initial $\bar{\nu}_e$ flux and the $\nu_\tau'$ flux. As these states propagate, the $\nu_2$ crosses adiabatic MSW-H and MSW-L resonances, therefore $F_{\nu_2}$ remains unchanged, and finally exits the SN as
\begin{equation}
F_{\nu_2} = (1-p_H) F_{\bar{\nu}_e}^0 + p_H F_{\nu_\tau'}^0\,.
\end{equation}
Meanwhile, after the RSFP-H, the $\nu_1$ flux can be written as $F_{\nu_1} = p_H F_{\bar{\nu}_e}^0+(1-p_H) F_{\nu_\tau'}^0
$. As it propagates outwards, it encounters two resonances as well, the non-adiabatic RSFP-L, and the adiabatic MSW-L (check inset plot).  At the RSFP-L point, transition happens between $\nu_1$ and $\bar{\nu}_1$. Since initial $\bar{\nu}_1$ flux deep inside the SN is $F_{\bar{\nu}_1}=F_{\nu_\mu'}^0$, after encountering the RSFP-L resonance, one has
\begin{eqnarray}
F_{\nu_1} &=& F_{\nu_\mu'}^0\,,\\
F_{\bar{\nu}_1} &=& p_H F_{\bar{\nu}_e}^0+(1-p_H) F_{\nu_\tau'}^0.
\end{eqnarray}
Finally, as the $\bar{\nu}_3$ is not affected by any of the resonances, we have
\begin{equation}
    F_{\bar{\nu}_3}=F_{\bar{\nu}_\tau'}^0.
\end{equation}
Therefore, the $\nu_e$ and the $\bar{\nu}_e$ fluxes at the Earth are
\begin{eqnarray}
    F_{\nu_e}&=&|U_{e1}|^2 F_{\nu_\mu'}^0 + |U_{e2}|^2 \left[(1-p_H) F_{\bar{\nu}_e}^0 + p_H F_{\nu_\tau'}^0\right]+ |U_{e3}|^2 F_{\nu_e}^0, \\
    F_{\bar{\nu}_e} &=& |U_{e1}|^2 \left[p_H F_{\bar{\nu}_e}^0+(1-p_H) F_{\nu_\tau'}^0 \right] + |U_{e2}|^2 F_{\bar{\nu}_\mu'}^0 + |U_{e3}|^2 F_{\bar{\nu}_\tau'}^0.
    \label{eq:NONeutrinoflux}
\end{eqnarray}
Now, if the neutrino magnetic moment were zero, there would be no RSFP, and the flux arriving at the Earth can be obtained by setting $p_H = 1$ in the above equations. Thus, in the limit where all the non-electron neutrino fluxes are set equal $(F_{\nu_\mu'}=F_{\nu_\tau'}=F_{\bar{\nu}_\mu'}=F_{\bar{\nu}_\tau'} = F_{\nu_x})$, one can write Eq.\,\ref{eq:NONeutrinoflux} as 
\begin{eqnarray}
    F_{\nu_e} &=& F_{\nu_e}^{MSW}-|U_{e2}|^2 (1-p_H)(F_{\nu_x}^0-F_{\bar{\nu}_e}^0) \\
    F_{\bar{\nu}_e}&=&F_{\bar{\nu}_e}^{MSW}+|U_{e1}|^2 (1-p_H)(F_{\nu_x}^0-F_{\bar{\nu}_e}^0)
    \label{Fnue-NH}
\end{eqnarray}
where $F_{\nu}^{MSW}= F_{\nu}(p_H=1) $.

With this information, we can calculate the changes in the neutronization fluence due to the presence of a non-zero TMM. From Fig.~\ref{fig:Neut_initial_flux}, note that for
$E\lesssim 20$ MeV, $F_{\bar{\nu}_e}^0 \approx F_{\nu_x}^0$, and hence the effect due to a non-zero TMM cancels out. On the other hand, for $E \gtrsim 20$ MeV, we have $F_{\nu_x}^0 > F_{\nu_e}^0$. This ends up reducing the $\nu_e$ flux, and increasing the $\bar{\nu}_e$ flux received at the Earth. 
We estimate that in case of complete adiabaticity of the RSFP-H resonance ($p_H=0$), for $E=25$ MeV, one has
\begin{equation}\label{ratios-NH}
    \frac{F_{\nu_e}}{F_{\nu_e}^{MSW}} \sim 0.85 \hspace{0.5cm} \text{and} \hspace{0.5cm}  \frac{F_{\bar{\nu}_e}}{F_{\bar{\nu}_e}^{MSW}} \sim 1.5 \hspace{0.5cm}
\end{equation}
Therefore, fingerprints of RSFP conversion can be visible in upcoming experiments sensitive to the neutronization burst flux of a galactic SN. 
\subsubsection{Inverted mass ordering (IO)}
A similar argument can be applied if the mass ordering is inverted (see bottom panel of Fig.\,\ref{fig:LevCros}). 
At very high densities, we have $F_{\nu_1}=F_{\nu_e}^0$ and $F_{\nu_2}=F_{\bar{\nu}_\tau'}^0$.
As the neutrinos propagate, a RSFP-H causes a flip between a $\nu_1$ and $\nu_2$ with probability $p_H$. Since the MSW-L resonance is adiabatic, the $\nu_2$ finally emerges as 
\begin{equation}
  F_{\nu_2}=p_H F_{\nu_e}^0 + (1-p_H) F_{\bar{\nu}_\tau'}^0.  
\end{equation}
On the other hand, the $\nu_1$, with flux $F_{\nu_1}=(1-p_H) F_{\nu_e}^0 + p_H F_{\bar{\nu}_\tau'}^0$ just after RSFP-H, goes through the adiabatic MSW-H and the non-adiabatic RSFP-E, where it flips completely to $\bar{\nu}_1$, and vice-versa. To get the final $\nu_1$ flux, we need to track the flavour evolution of $\bar{\nu}_1$.
The $\bar{\nu}_1$ is produced deep inside as a $\bar{\nu}_e$, while $\bar{\nu}_3$ is produced as a $\nu_\mu'$. At RSFP-L, $\bar{\nu}_1$ and $\bar{\nu}_3$ exchange fluxes so that $F_{\bar{\nu}_1}=F_{\nu_\mu'}^0$ and $F_{\bar{\nu}_3}=F_{\bar{\nu}_e}^0$. The $\bar{\nu}_3$ flux remains unchanged and exits the SN as
\begin{equation}
   F_{\bar{\nu}_3}=F_{\bar{\nu}_e}^0.
\end{equation}
On the other hand, the $\bar{\nu}_1$ exchanges flux with $\nu_1$ at RSFP-E and reaches the surface of the star as
\begin{equation}
    F_{\bar{\nu}_1}=(1-p_H) F_{\nu_e}^0 + p_H F_{\bar{\nu}_\tau'}^0\,,
\end{equation}
whereas the $\nu_1$ is emitted as 
\begin{equation}
    F_{\nu_1}=F_{\nu_\mu'}^0\,.
\end{equation}
The $\bar{\nu}_2$ and $\nu_3$ have relatively trivial dynamics, and they are emitted as 
\begin{eqnarray}
  F_{\nu_3} &=& F_{\nu_\tau'}^0\,,\\
  F_{\bar{\nu}_2} &=& F_{\bar{\nu}_\mu'}^0.
\end{eqnarray}
Collecting these fluxes, the $\nu_e$ and $\bar{\nu}_e$ fluxes at the Earth are given by
\begin{eqnarray}\label{eq:NuefluxIO}
  F_{\nu_e}&=&|U_{e1}|^2 F_{\nu_\mu'}^0 + |U_{e2}|^2 \left[p_H F_{\nu_e}^0 + (1-p_H) F_{\bar{\nu}_\tau'}^0\right]+ |U_{e3}|^2 F_{\nu_\tau'}^0\,,\\
  \label{eq:NuebarfluxIO}
  F_{\bar{\nu}_e} &=& |U_{e1}|^2 \left[(1-p_H) F_{\nu_e}^0 + p_H F_{\bar{\nu}_\tau'}^0 \right] + |U_{e2}|^2 F_{\bar{\nu}_\mu'}^0 + |U_{e3}|^2 F_{\bar{\nu}_e}^0\,.
\end{eqnarray}
In terms of the pure MSW fluxes (which is obtained by setting $p_H=1$ in Eqs.\,(\ref{eq:NuefluxIO}-\ref{eq:NuebarfluxIO})), we get
\begin{eqnarray}
    F_{\nu_e}&=&F_{\nu_e}^{MSW}-|U_{e2}|^2 (1-p_H)(F_{\nu_e}^0- F_{\nu_x}^0)\,\\
    F_{\bar{\nu}_e}&=&F_{\bar{\nu}_e}^{MSW}+|U_{e1}|^2 (1-p_H)(F_{\nu_e}^0- F_{\nu_x}^0)\,,
\end{eqnarray}
where all the non-electron neutrino fluxes are set equal $(F_{\nu_\mu'}=F_{\nu_\tau'}=F_{\bar{\nu}_\mu'}=F_{\bar{\nu}_\tau'} = F_{\nu_x})$.
Thus, the RSFP signal for inverted ordering will be more relevant at $E\lesssim 20$ MeV once $F_{\nu_e}^0 \gg F_{\nu_x}^0$. At $E=10$ MeV we estimate 

\begin{equation} \label{ratios-IH}
    \frac{F_{\nu_e}}{F_{\nu_e}^{MSW}} \sim 0.3 \hspace{0.5cm} \text{and} \hspace{0.5cm}  \frac{F_{\bar{\nu}_e}}{F_{\bar{\nu}_e}^{MSW}} \sim 7.
\end{equation}
From Eqs.~(\ref{ratios-NH}) and (\ref{ratios-IH}), the RSFP signal will be more dominant in the IO compared to NO, due to a large enhancement in the $\bar{\nu}_e$ flux, which is quite uncharacteristic of the neutronization epoch. As a result, experiments sensitive to $\bar{\nu}_e$ from a SN will suddenly see a large flux during the shock breakout phase, which can be a smoking-gun signal of the presence of neutrino TMM.
\section{Sensitivity in upcoming experiments}\label{SEC-05}
In this section, we study the impact of RSFP in the neutronization burst signal at upcoming neutrino detectors.
As discussed in the previous sections, the effect of the RSFP is to reduce the $\nu_e$ flux, and increase the $\bar{\nu}_e$. Depending on the mass ordering, the effect can be quite large. As a result, sensitivity can be maximised using a combination of detectors sensitive to the $\nu_e$ and the $\bar{\nu}_e$ flux. We focus on the Deep Underground Neutrino Experiment (DUNE), which can detect $\nu_e$, and the Hyper-Kamiokande (HK) which is mainly sensitive to the $\bar{\nu}_e$ flux. The number of events of species $\nu_\alpha$ detected per unit energy is
\begin{equation} \label{event-spectrum}
    \frac{dN_{\nu_{\alpha}}}{dE_r}=\frac{N_{\rm tar}}{4 \pi R^2} \int d E_t  F_{\nu_{\alpha}}(E_t)\sigma_{\alpha}(E_t) W(E_r,E_t),
\end{equation}
with $N_{\rm tar}$ the number of targets in the detector material, $R$ the distance of the supernova from the Earth, $F_{\nu_\alpha}$ is the $\nu_\alpha$ flux that reaches the detector, $\sigma_\alpha$ is the relevant neutrino interaction cross section and $W$ is the Gaussian energy resolution function with width $\sigma_E$ that depends on that experimental setup. $E_t$ is the actual value of neutrino energy, while $E_r$ is the experimentally reconstructed energy. 
For the analysis presented in this paper, we choose a fiducial distance of $R=10\,$kpc. For the neutrino oscillation parameters, we assume the best fit values in \cite{deSalas:2020pgw}.

The TMM matrix $M$ in Eq.\,\ref{M-matrix}, with $\mu_{\mu \tau}=0$, has two independent elements $\mu_{e \mu}$, $\mu_{e \tau}$.  We consider the following cases:
\begin{enumerate}
    \item \label{case 1} both $\mu_{e \mu}$ and $\mu_{e \tau}$ are equal, set to $\mu_\nu$, and considered free parameters. 
    
    \item \label{case 2} $\mu_{e \mu}=0$ and $\mu_{e \tau}=\mu_\nu$ is a free parameter.
    
    \item \label{case 3} $\mu_{e \mu}=\mu_\nu$ is considered a free parameter, while $\mu_{e \tau}=0$.
\end{enumerate}
Because the TMMs always appear together with $B_\perp (r)$ in the equations, all our results are sensitive to the quantity $\mu_\nu B_0$, and not independently to $\mu_\nu$ or $B_0$. Hence, to derive constraints on $\mu_\nu$, one needs prior knowledge of $B_0$ from simulations and/or astrophysical observations.

The schemes of level crossing and resonances are the same for all cases. For the specific interval in eq.~(\ref{muB-range}), MSW-H and MSW-L are always adiabatic while RSFP-L and RSFP-E are non-adiabatic. However, the degree of adiabaticity of RSFP-H, given by the hoping probability $p_H$, is strongly dependent on the configuration of the matrix $M$, specifically on the magnitude of $\mu_{e \tau'}$ (see Eq.~(\ref{ad-RSFP})). For the set of mixing parameters and magnetic field assumed in this work, $(\mu_{e \tau'})_1>(\mu_{e \tau'})_3>(\mu_{e \tau'})_2$ for the entire energy range close to RSFP-H, where the subscript points to one of the cases above. Therefore, the strength of conversions at RSFP-H has the following hierarchy:  case \ref{case 1}$>$ case \ref{case 3} $>$ case \ref{case 2} for the same value of $\mu_{\nu} B_0$. As a consequence, the fluxes on Earth will change accordingly and that translates into different experimental sensitivities.

The DUNE underground facility in South Dakota uses $40$-kton of liquid argon in time projection chambers. The experiment will be able to detect neutrinos from few MeV up to GeV. The relevant interaction for SN neutrinos that reach the detector is the charged current scattering with argon nuclei, $\nu_{e}+{ }^{40}\mathrm{Ar} \rightarrow { }^{40}\mathrm{K}^{*}+e^{-}$.
The final state consists of electrons, and excited states of the potassium nuclei, which de-excite producing a cascade of photons in the energy range between 5 and 50 MeV. For our analysis, we use the Monte Carlo event simulator MARLEY to generate the $\nu_e-\mathrm{Ar}$ interaction cross section. The energy resolution is taken to be $\sigma_E/\mathrm{MeV}=0.11 \sqrt{E_{r} / \mathrm{MeV}}+0.02 E_{r} / \mathrm{MeV}$, which implies that $\sigma_E/E_r \sim 5 \%$ at $10$ MeV. In addition, we use bins of $5\,$MeV, with the energy reconstruction threshold set at $4\,$MeV.

HK, an upgrade of the Super-Kamiokande (SK) detector, is a next generation water-Cherenkov detector, with a fiducial volume of $187$ kt in each of its two tanks. It is expected to record a huge number of events in case of a galactic SN. HK primarily detects $\bar{\nu}_e$ through the Inverse Beta Decay (IBD) channel, $ \bar{\nu}_{e}+p \rightarrow e^{+}+n$. The IBD cross section for the MeV range is extracted from \cite{Strumia:2003zx} and the energy resolution is the same of SK, namely $\sigma_E/\mathrm{MeV}=0.6 \sqrt{E_r/\mathrm{MeV}}$ with $\sigma_E/E_r \sim 20 \%$ at $10$ MeV. We simulate events at HK detector using bins of $8$ MeV, and a detector threshold of $3$ MeV.

\begin{figure}[!t]
\centering
  \includegraphics[width=0.45\textwidth]{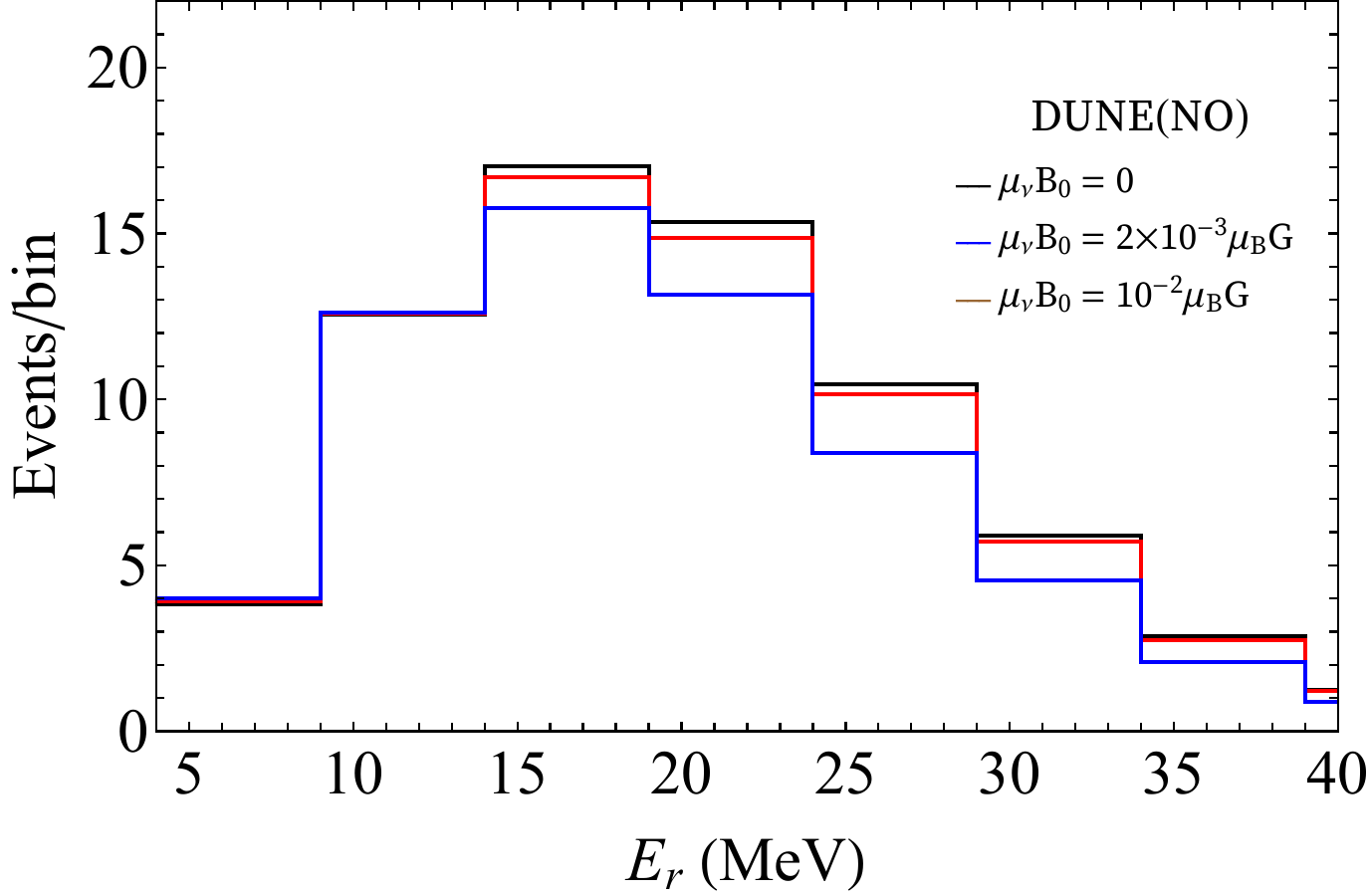}~~
  \includegraphics[width=0.45\textwidth,height=0.3\textwidth]{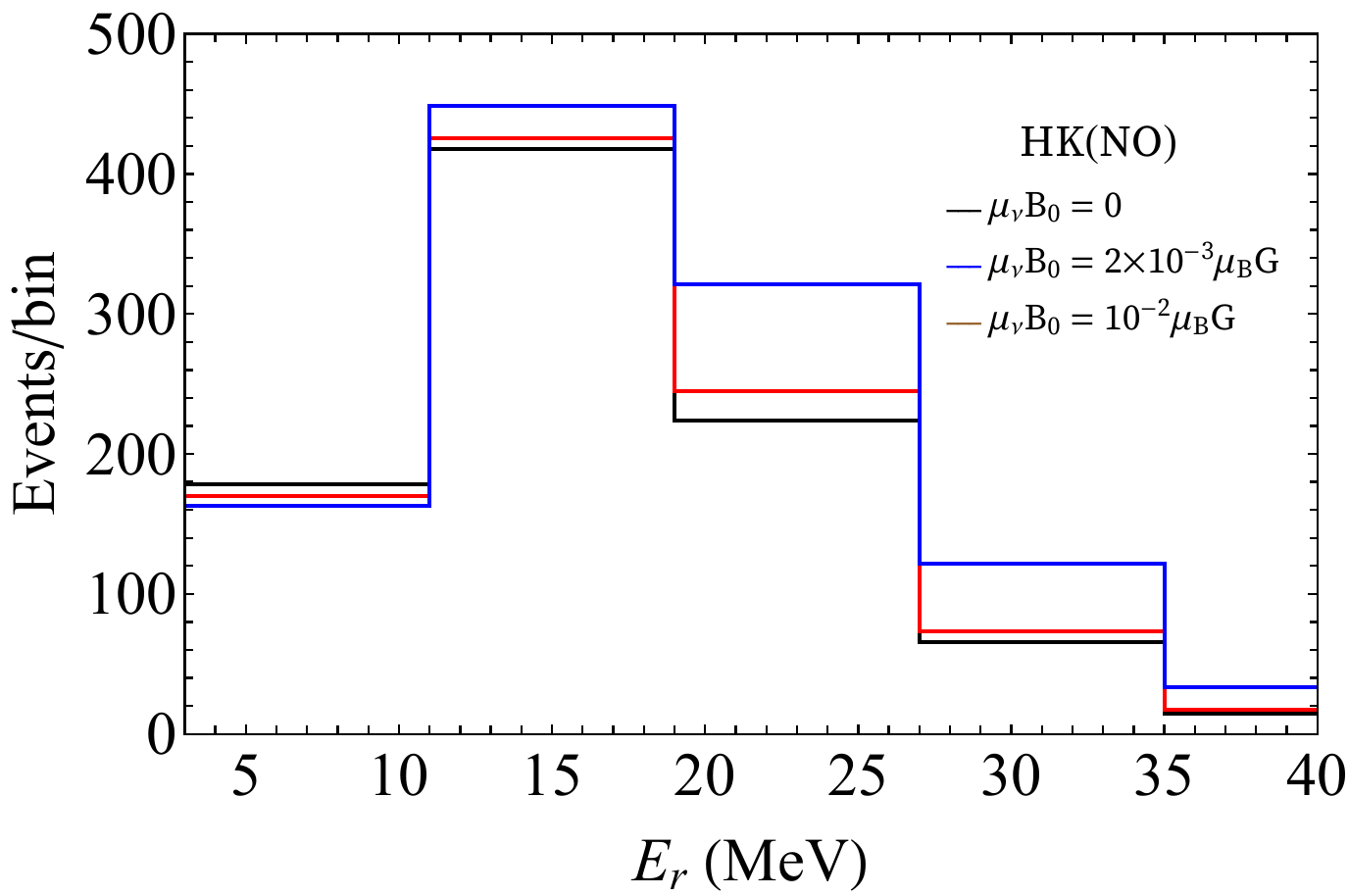}
  \caption{Expected event spectrum for a supernova explosion $10$ kpc away in DUNE (left, $\nu_e$ channel) and HK (right, $\bar{\nu}_e$ channel) for distinct values of $\mu_\nu B_0$. We assume NO for neutrino masses. }
\label{spectrum-NO}
\end{figure}

We compute the number of events at each detector for values of the parameter $\mu_\nu B_0$ lying within the range given in (\ref{muB-range}), such that the RSFP-L/E are non-adiabatic, whereas the RSFP-H can transition from non-adiabatic to adiabatic. We focus on the most general Case \ref{case 1}. In Fig.~\ref{spectrum-NO}, we show how the neutrino event spectra is altered at DUNE and HK, assuming normal mass ordering, in the presence of a non-zero neutrino TMM. The figure nicely illustrates the features we discussed in Sec.~\ref{level-crossing} using the level crossing diagram: a spectral distortion, resulting in  suppression of events for DUNE, while simultaneously enhancing the events in HK, for bins higher than $10$ MeV. We find that our results are sensitive to values of $\mu_\nu B_0$ as small as $10^{-3} \mu_B\,$G; the spectral distortion becomes more prominent as these values are increased. Case \ref{case 2} and \ref{case 3} also demonstrate similar spectral distortions, albeit smaller, and hence we do not show them here.

\begin{figure}[!t]
  \includegraphics[width=0.45\textwidth]{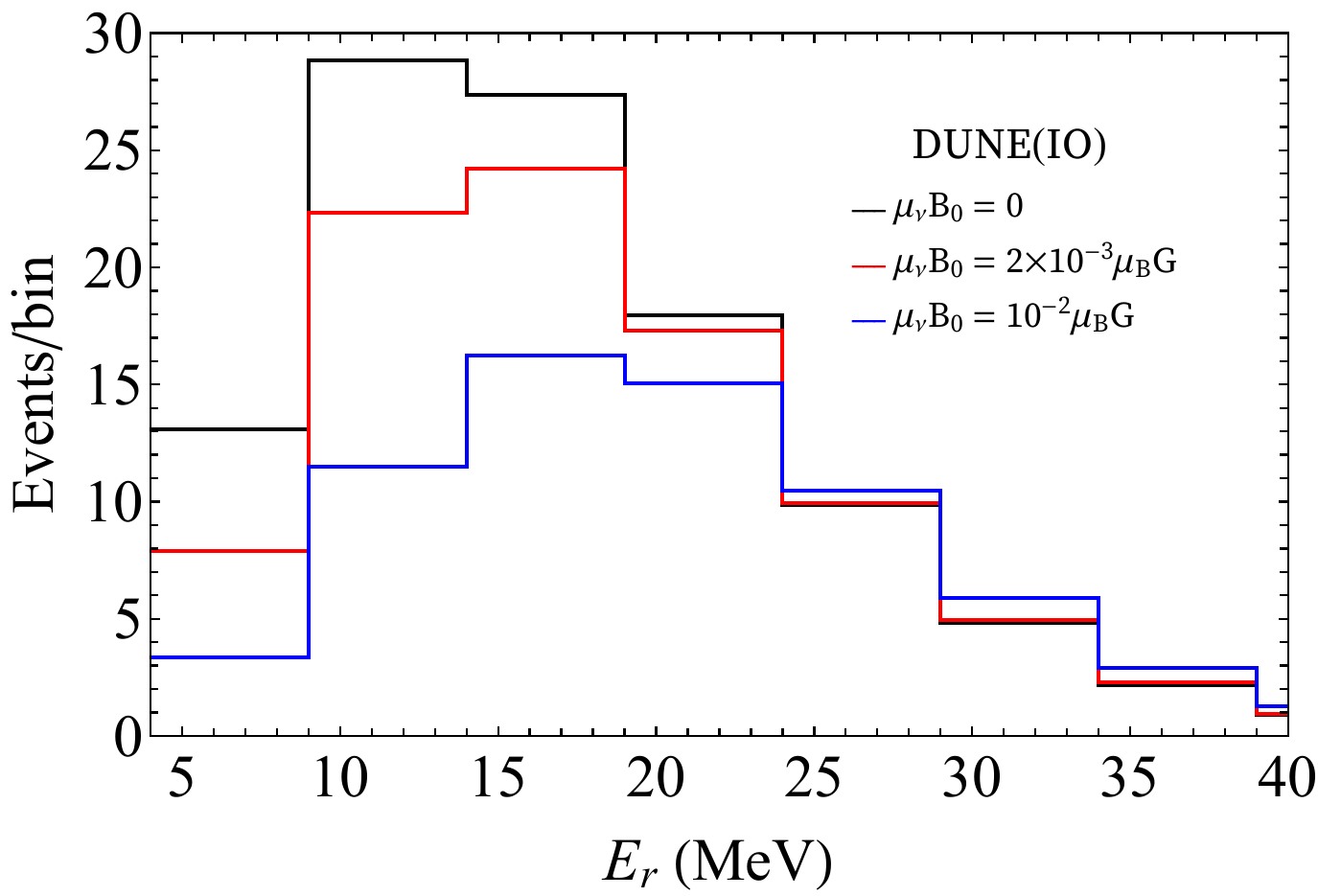}~~
  \includegraphics[width=0.455\textwidth, height=0.295\textwidth]{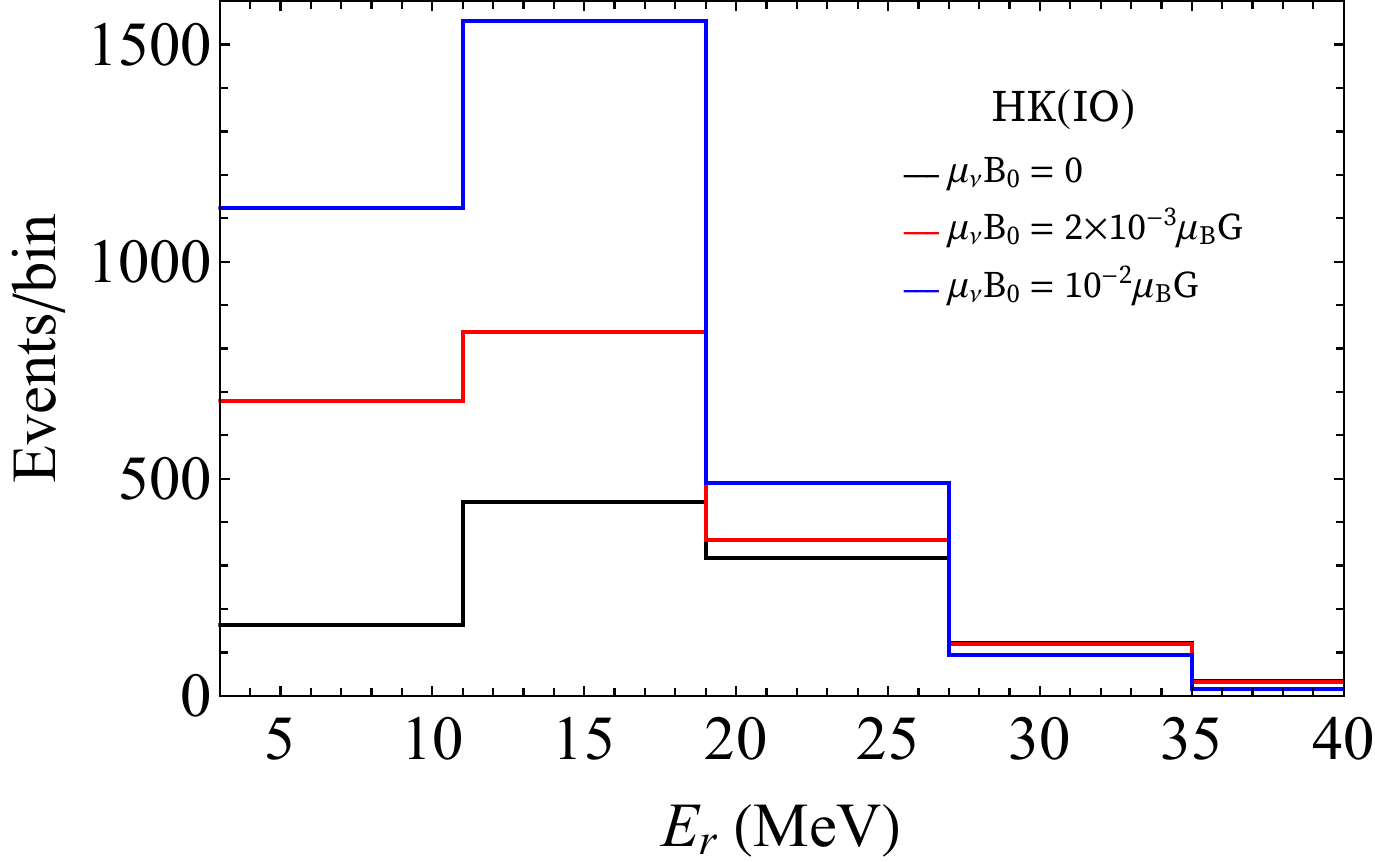}
  \caption{Expected event spectrum for a supernova explosion $10$ kpc away in DUNE (left, $\nu_e$ channel) and HK (right, $\bar{\nu}_e$ channel) for distinct values of $\mu_\nu B_0$. We assume IO for neutrino masses.  }
\label{spectrum-IO}
\end{figure}

In fig.~\ref{spectrum-IO}, we show a similar analysis for the IO, where this effect is expected to be more drastic. We find that for energy bins $E_r \lesssim 25$ MeV, there is a sharp decrease in events for DUNE, while HK records an almost three-fold increase in number of events. This is consistent with our understanding in Sec.~\ref{level-crossing} (see the estimations in eq.~(\ref{ratios-IH})). The decrease in events in DUNE, combined with the sharp increase in HK, illustrates that the combination of RSFP-H with MSW-H is very efficient in converting $\nu_e$ to $\bar{\nu}_e$. This can be considered as a smoking gun signal of non-zero TMMs of neutrinos.


We perform a $\chi^2$ analysis to study the sensitivity of DUNE and HK to the presence of neutrino TMM, using the signals from a future galactic SN. We use the Poissonian $\chi^2$ estimator,
\begin{equation} \label{chisq}
    \chi^{2}=\sum_{i=1}^{n} 2\left[F_{i}-D_{i}+D_{i} \ln \left(D_{i} / F_{i}\right)\right]\,,
\end{equation}
where $F_i$ is the number of events predicted for a specific non zero value of $\mu_\nu B_0$ and $D_i$ is the expected number of events assuming $\mu_\nu B_0=0$. The subscript $i$ refers to the $i$-th bin and $n$ is the number of bins of the experiment.

\begin{figure}[!t]
\centering
  \includegraphics[width=0.5\textwidth]{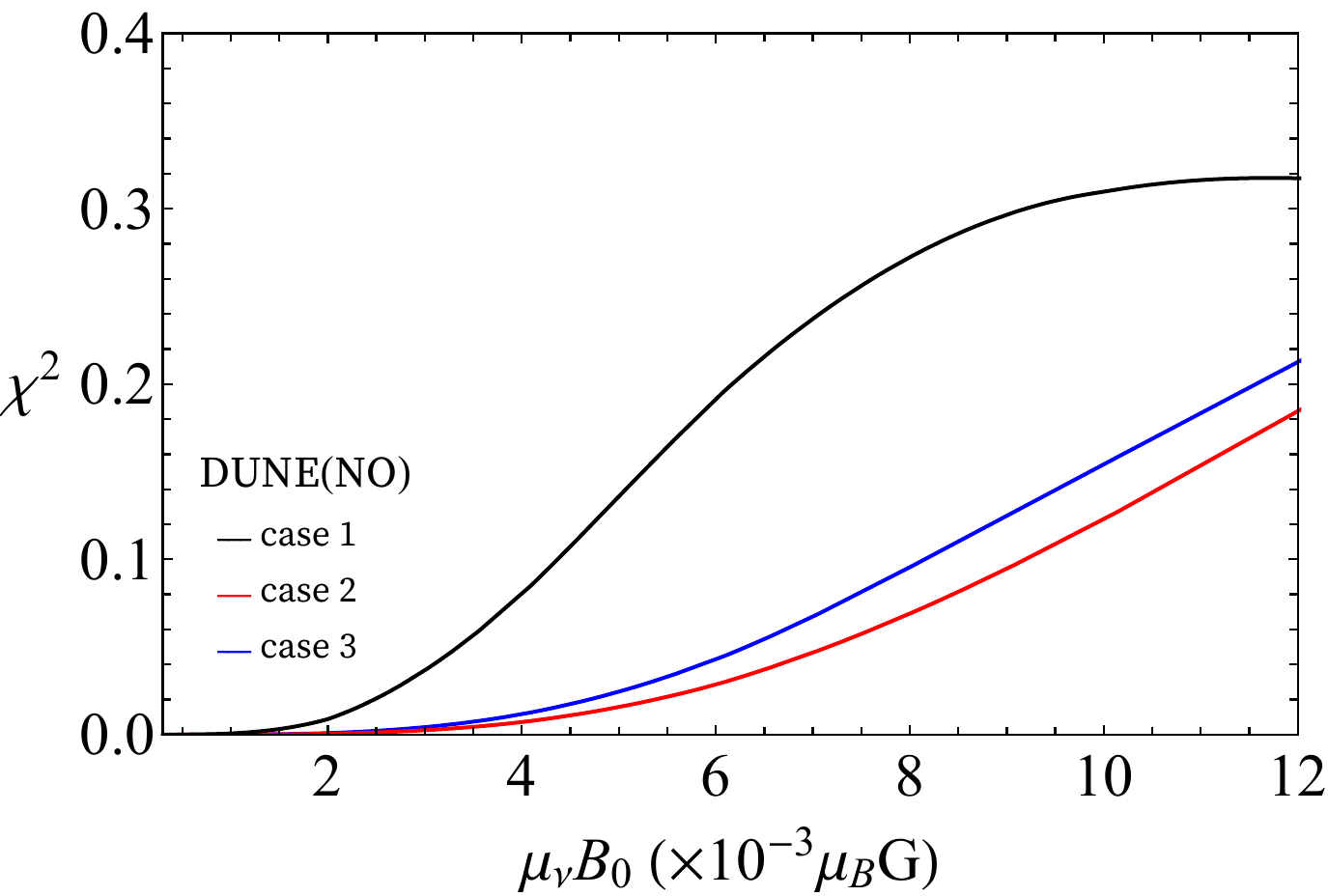}~
  \includegraphics[width=0.5\textwidth,height=0.325\textwidth]{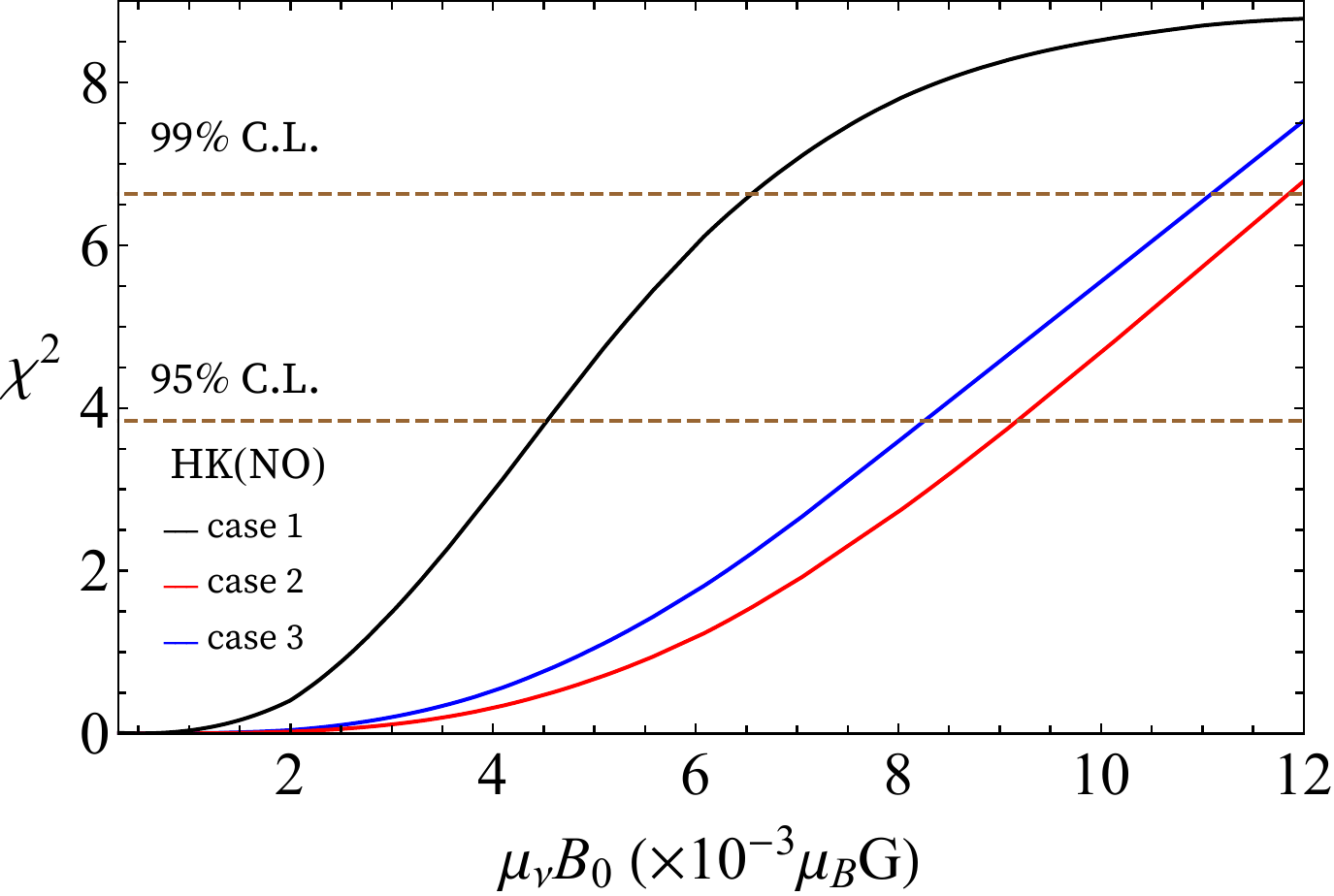}
  \caption{$\chi^2$ in Eq.\,(\ref{chisq}) as a function of $\mu_\nu B_0$ assuming NO for DUNE (left) and HK (right). The analysis compares the RSFP hypothesis with the pure MSW one for the different cases \ref{case 1}, \ref{case 2} and \ref{case 3} defined in the text. As expected, the strength of the signal and hence the sensitivity is always greater for case \ref{case 1} for both mass orderings.} 
  \label{chisq-NO}
\end{figure}

\begin{figure}[!t]
  \includegraphics[width=0.5\textwidth]{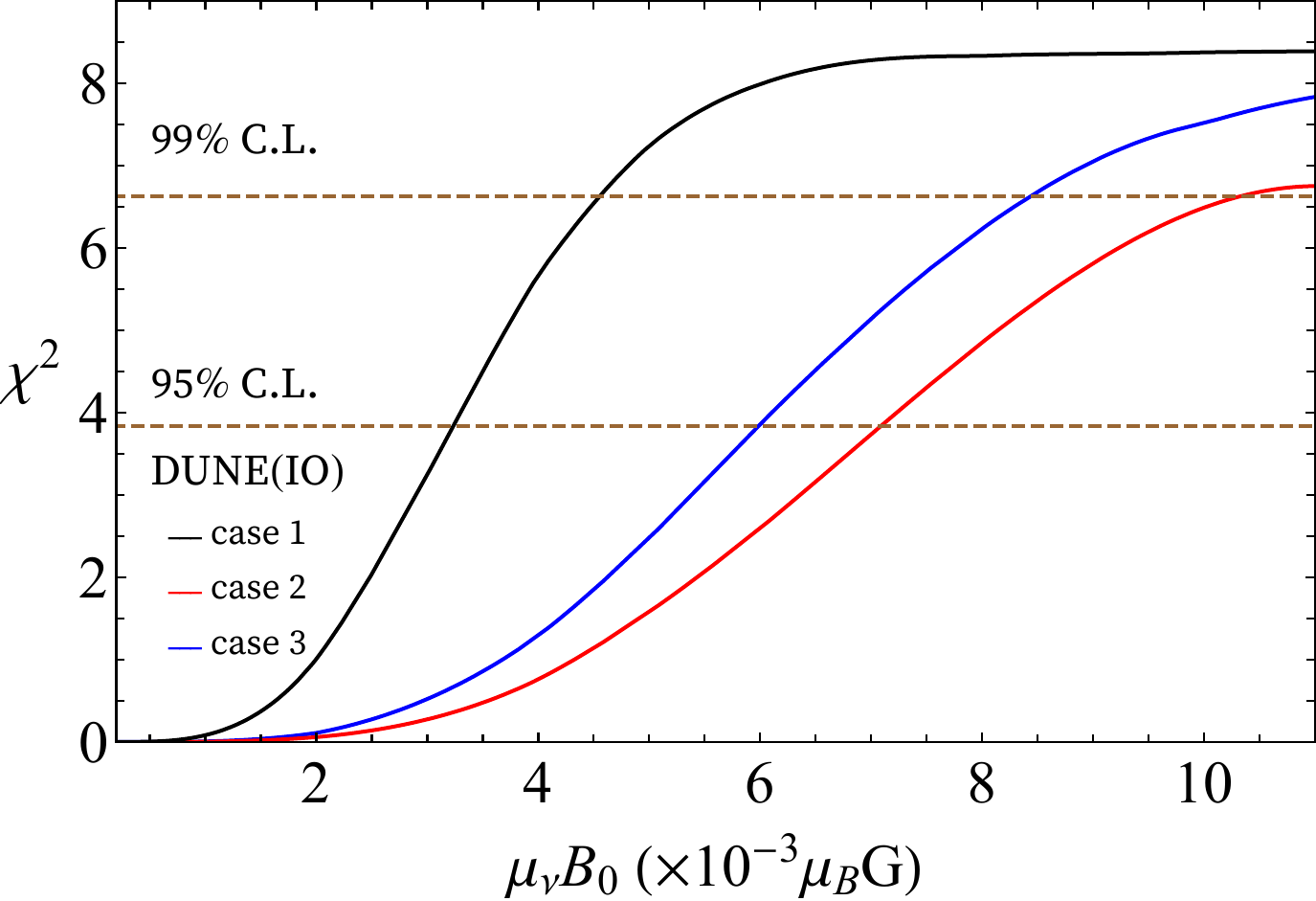}~
  \includegraphics[width=0.5\textwidth]{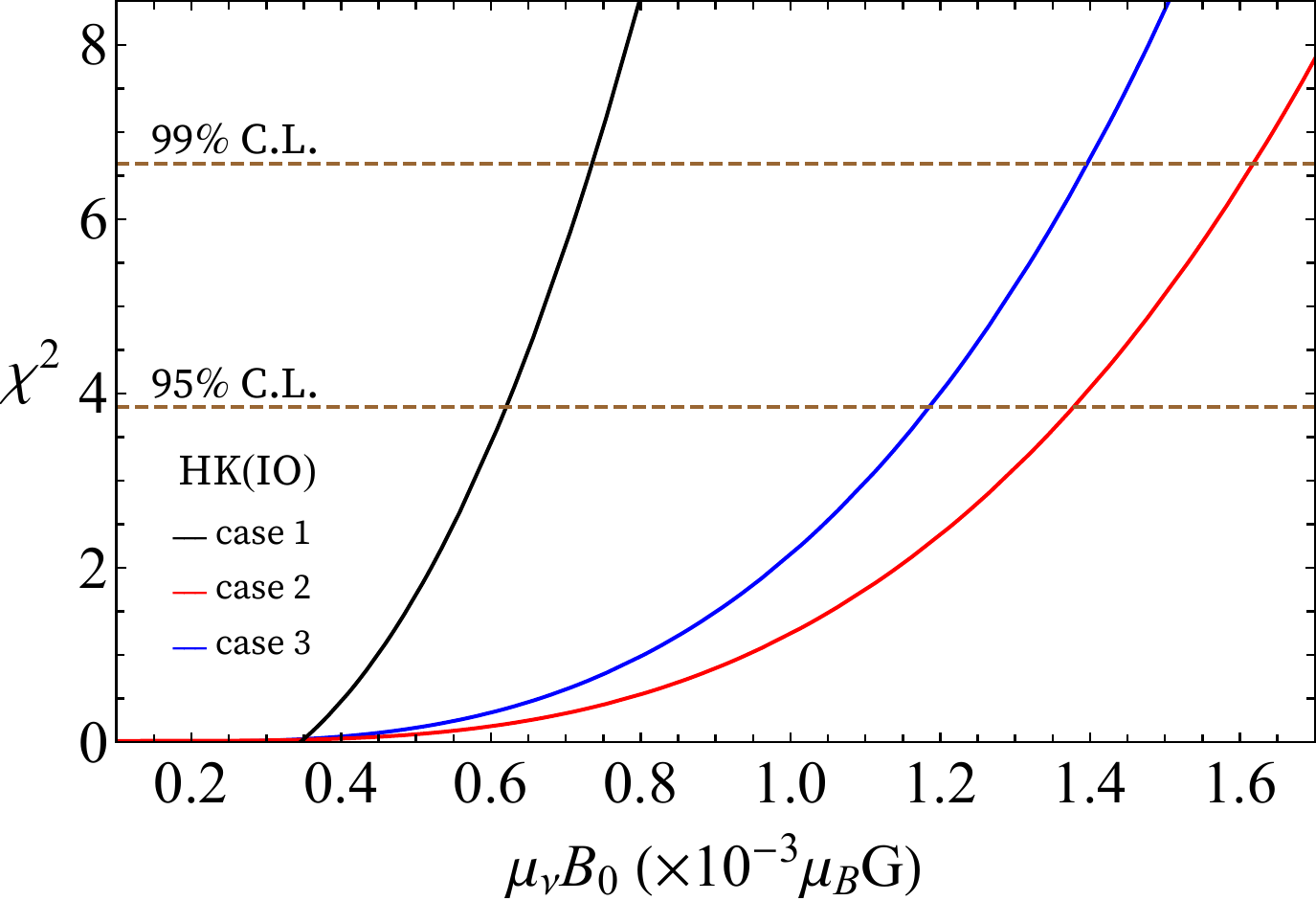}
  \caption{$\chi^2$ in Eq.~(\ref{chisq}) as a function of $\mu_\nu B_0$ assuming NO for DUNE (left) and HK (right) that compares the RSFP hypothesis with the pure MSW one for the different cases \ref{case 1}, \ref{case 2} and \ref{case 3} defined in the text. }
\label{chisq-IO}
\end{figure}

Figs.~\ref{chisq-NO} and \ref{chisq-IO} show $\chi^2$ as a function of $\mu_\nu B_0$ for NO and IO, respectively. The null hypothesis is that neutrinos undergo pure MSW flavour conversions, and there is no RSFP involved. As discussed before, the results are sensitive to the magnitude of the product $\mu_\nu B_0$, and it is not possible to disentangle one effect from the other. DUNE shows very little sensitivity to RSFP for NO, as we show in Fig.~\ref{chisq-NO}. On the other hand, for IO (see Fig.\,\ref{chisq-IO}), DUNE can exclude at $95 \%$ C.L. ,
\begin{eqnarray} \label{limits-DUNE-IO}
    (\mu_\nu B_0)_1 &\gtrsim& 3 \times 10^{-3} \mu_B \text{G}\,,\\ (\mu_\nu B_0)_2 &\gtrsim& 7 \times 10^{-3} \mu_B \text{G}\,,\\ (\mu_\nu B_0)_3 &\gtrsim& 6 \times 10^{-3}\mu_B \text{G}\,,
\end{eqnarray}
where the subscripts refers to specific cases \ref{case 1}, \ref{case 2} and \ref{case 3} respectively.

Due to its large size, HK will be sensitive to both the mass orderings. For the NO, HK can exclude, at $95 \%$ C.L.,
\begin{eqnarray} \label{limits-HK-NO}
    (\mu_\nu B_0)_1 &\gtrsim& 4.5 \times 10^{-3} \mu_B \text{G}\,,\\  (\mu_\nu B_0)_2 &\gtrsim& 9 \times 10^{-3} \mu_B \text{G}\,,\\  (\mu_\nu B_0)_3 &\gtrsim& 8 \times 10^{-3}\mu_B \text{G}\,.
\end{eqnarray}
The best sensitivity to TMM comes from HK in case of IO, and it can exclude, at $95 \%$ C.L.,
\begin{eqnarray} \label{limits-HK-IO}
    (\mu_\nu B_0)_1 &\gtrsim& 0.6 \times 10^{-3} \mu_B \text{G}\,,\\
    (\mu_\nu B_0)_2 &\gtrsim& 1.4 \times 10^{-3} \mu_B \text{G}\,,\\
    (\mu_\nu B_0)_3 &\gtrsim& 1.2 \times 10^{-3}\mu_B \text{G}. \label{limits-HK-IO-3}
\end{eqnarray}
As discussed, for a given $B_0$, these bounds above can be translated into bounds on the TMM of neutrinos. Table~\ref{summary_NMM} summarizes the derived sensitivities to $\mu_\nu$ assuming $B_0=10^{10}\,$G.

\begin{table}[] 
\centering 
\resizebox{0.8\textwidth}{!}{%
\begin{tabular}{|c|llllll|} 
\hline
\hline
\multirow{3}{*}{\textbf{Experiments}} &
  \multicolumn{6}{l|}{\textbf{Sensitivities on Neutrino Magnetic Moments (in $\mu_B$)}} \\ \cline{2-7} 
 &
  \multicolumn{2}{c|}{\textbf{CASE 1}} &
  \multicolumn{2}{c|}{\textbf{CASE 2}} &
  \multicolumn{2}{c|}{\textbf{CASE 3}} \\ \cline{2-7} 
 &
  \multicolumn{1}{c|}{\textbf{NO}} &
  \multicolumn{1}{c|}{\textbf{IO}} &
  \multicolumn{1}{c|}{\textbf{NO}} &
  \multicolumn{1}{c|}{\textbf{IO}} &
  \multicolumn{1}{c|}{\textbf{NO}} &
  \multicolumn{1}{c|}{\textbf{IO}} \\ \hline
\textbf{HK} &
  \multicolumn{1}{l|}{$4.5 \times 10^{-13}$} &
  \multicolumn{1}{l|}{$6 \times 10^{-14}$} &
  \multicolumn{1}{l|}{$9 \times 10^{-13}$} &
  \multicolumn{1}{l|}{$1.4 \times 10^{-13}$} &
  \multicolumn{1}{l|}{$8 \times 10^{-13}$} &
  \multicolumn{1}{l|}{$1.2 \times 10^{-13}$} \\
   \hline
\textbf{DUNE} &
  \multicolumn{1}{l|}{$-$} &
  \multicolumn{1}{l|}{$3 \times 10^{-13}$} &
  \multicolumn{1}{l|}{$-$} &
  \multicolumn{1}{l|}{$7 \times 10^{-13}$} &
  \multicolumn{1}{l|}{$-$} &
  \multicolumn{1}{l|}{$6 \times 10^{-13}$} \\
  \hline
  \hline 
\end{tabular}
}
\caption{Experimental sensitivities on neutrino magnetic moments for different benchmark scenarios for a fixed value of magnetic field strength $B_0 = 10^{10}$ G.}
\label{summary_NMM}
\end{table}

Uncertainties may affect the derived sensitivity. However, such impact need not be drastic as the neutronization burst is a robust feature of all SN simulations with a relatively weak dependence on progenitor mass and/or specific traits of the models \cite{Kachelriess:2004ds, Serpico:2011ir, OConnor:2018sti, Tang:2020pkp}. Indeed, other studies show that different progenitor models and neutrino emission parameters yield changes that are smaller than the statistical uncertainties for current water Cherenkov detectors \cite{Kachelriess:2004ds}. For this reason, we do not perform a dedicated study about the influence of uncertainty in the fluence parameters on the expected sensitivities to $\mu_\nu B_0$.

Systematic uncertainties can also impact the above sensitivities, but owing to energy dependent spectral distortion of the neutrino signal in case of nonzero $\mu_\nu B_0$, only mild changes are expected. To see how systematics affect the $\chi^2$, we modify Eq.~(\ref{chisq}) as
\begin{equation} \label{chisq-sys}
    \chi^{2}=\sum_{i=1}^{n} 2\left[(1+\xi)F_{i}-D_{i}+D_{i} \ln \left(\frac{D_{i} }{(1+\xi) F_{i}}\right)\right]+\frac{\xi^2}{\sigma^2} \,,
\end{equation}
where $\xi$ is nuisance parameter and $\sigma$ an overall normalization error that we assume to be $50 \%$. $\xi$ is allowed to vary in the $3 \sigma$ range and we minimize the $\chi^2$ over it. We focus on case \ref{case 1} and summarize new exclusion limits at $95 \%$ C.L. below
\begin{eqnarray} \label{limits-sys}
  \text{DUNE(IO):} \hspace{0.2cm} (\mu_\nu B_0)_1 &\gtrsim& 3.5 \times 10^{-3} \mu_B \text{G}\,,\\
   \text{HK(NO):} \hspace{0.2cm} (\mu_\nu B_0)_1 &\gtrsim& 6 \times 10^{-3} \mu_B \text{G}\,,\\
   \text{HK(IO):} \hspace{0.2cm} (\mu_\nu B_0)_1 &\gtrsim& 0.7 \times 10^{-3}\mu_B \text{G},
\end{eqnarray}
with no sensitivity from DUNE in case of normal ordering.

A final comment on the adiabaticity of the RSFP-E resonance is in order. For $\mu_\nu B_0 > 10^{-2}$ $\mu_B \text{G}$, the RSFP-E starts to be partially adiabatic. This can end up affecting the energy spectra in both the mass orderings. We checked that it is possible for the event spectra in the NO even to mimic that of the IO in certain cases. We do not comment on these cases further and leave the analysis to a future study.

\section{Implications on neutrino properties}
In this section, we mull over the possible impact of our results on the uncharted territories of neutrino physics. To do so, we first briefly summarize the experimental constraints and theoretical predictions on neutrino magnetic moments, along with our results, in Fig.~\ref{sum}. Existing limits on neutrino magnetic moments are shown in blue, while the red lines indicate the future sensitivities on neutrino magnetic moments at DUNE and HK, based on our analysis, by setting magnetic field strength $B_0 =10^{10}$ G. Note that as compared to existing limits, the sensitivity can be upgraded by two or three orders of magnitude (from $\mathcal{O} [10^{-11}]~ \mu_B$ to $\mathcal{O} [10^{-14}]~ \mu_B$) by utilizing the neutrino spectra from the SN neutronization  burst phase in forthcoming neutrino experiments like the DUNE, and HK. This is more stringent than the expected sensitivity when considering the scattering at the DUNE near detector \cite{Mathur:2021trm}. This will have consequences for three of the most important unanswered questions in neutrino physics:
\begin{figure}[t!] 
  \centering
  \includegraphics[width=0.9\textwidth]{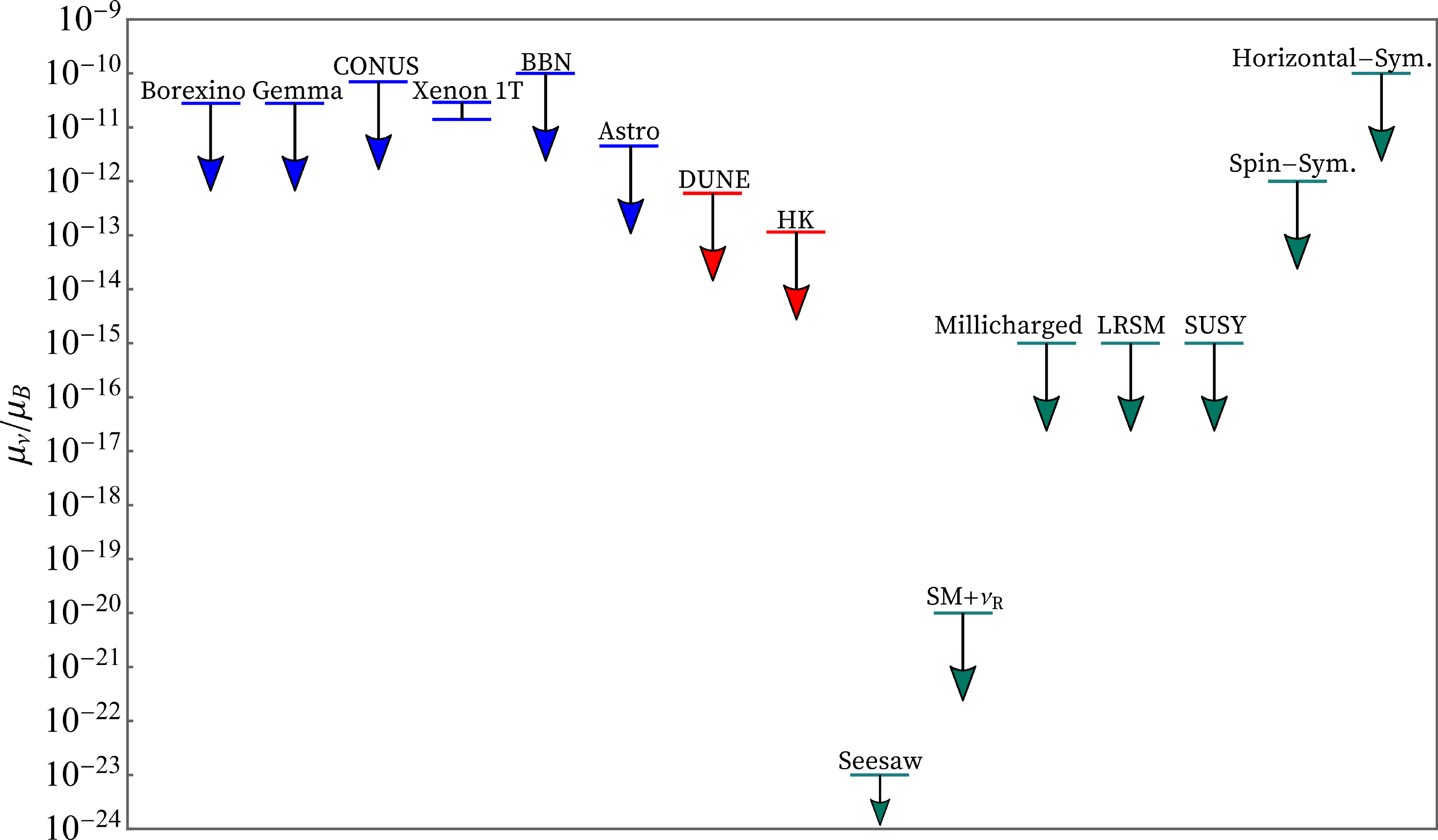}
  \caption{Summary of experimental constraints and theoretical predictions on neutrino magnetic moments. The red lines indicate the future sensitivities on neutrino magnetic moments at DUNE and HK experiments based on our analysis. Here we set magnetic field strength $B_0 =10^{10}$ G. Theoretical predictions for maximum possible strengths of neutrino  magnetic moments are summarized from Ref.~\cite{Babu:2020ivd} and references therein. See text for details.}
  \label{sum}
\end{figure}
\begin{itemize}
    \item {\textsl{Dirac/Majorana nature of neutrinos:} }
    Based on an Effective Field Theory (EFT) analysis, it has been shown that Dirac neutrino magnetic moments over $10^{-14}\mu_B$ would not be natural, since this would generate unacceptably large neutrino masses at higher loops \cite{Bell:2005kz}. Weak interaction corrections to neutrino mass originating from the magnetic moment operator are excessive in the case of a Dirac neutrino magnetic moment. However, such a correction is small for Majorana neutrinos since it is proportional to the mass differences of charged leptons. 
    Because of EFT naturalness considerations, the transition magnetic moments for Majorana neutrinos are permitted to be significantly greater than the Dirac scenario  \cite{Davidson:2005cs,Bell:2006wi}. For instance, if the new physics scale is roughly a TeV, EFT would allow the transition magnetic moment as large as  $\mu_{\nu} \sim 10^{-7}\mu_B$. Therefore, if DUNE or Hyper-Kamiokande experiment measure neutrino magnetic moments at the level of $10^{-13}~ \mu_B$, it is more likely that neutrinos are Majorana in nature.
    \item {\textsl{Neutrino mass ordering:}} The SN neutronization burst has long been advertised as one of the cleanest means to identify the neutrino mass ordering. The presence (absence) of a distinct peak during the burst phase indicates inverted (normal) mass ordering. However, this is true only in the absence of any new physics in the neutrino sector. If neutrinos have a finite magnetic moment, it can result in the suppression of the peak in the IO due to $\nu_e\rightarrow \bar{\nu}_e$ conversion. As a result, this might make identification of the mass ordering from the burst phase ambiguous. 
    \item {\textsl{Neutrino mass generation mechanism:}} Due to similar chiral structure of the neutrino magnetic moment and neutrino mass operator, while removing the photon line from the loop diagram that yields the neutrino magnetic moment, a neutrino mass term is generated, and $\mu_\nu$ generally becomes proportional to $m_\nu$. One can estimate $m_\nu$ originating from such diagrams as
\begin{equation}
    m_\nu \sim \frac{ \mu_\nu}{\mu_B} \,\frac{\Lambda^2}{2 m_e},
    \label{order}
\end{equation}
     where $\Lambda$ is the mass scale of the  heavy particle inside the loop diagram.  Experimental searches generally prefer charged particles with masses heavier than 100 GeV \cite{Babu:2019mfe}. Without extra symmetries (and in the absence of severe fine-tuning), this implies that neutrino mass $m_\nu$ must be $\sim 0.1$ MeV in order to generate neutrino magnetic moments in the order of $10^{-11}~\mu_B$ for a new physics scale $\sim 100$ GeV. This prediction of active neutrino mass contradicts the observed neutrino masses by six orders of magnitude. Therefore, one requires leptonic symmetries \cite{Voloshin:1987qy, Barr:1990um, Babu:2020ivd} in order to get large neutrino magnetic moments while being consistent with tiny neutrino masses and mixings utilizing their different Lorentz structure.
     There are several BSM extensions to generate neutrino masses and mixings. However, most of these extensions cannot accommodate large neutrino magnetic moments ($\gtrsim 10^{-14}\mu_B$). For instance, if right-handed neutrinos are introduced in the SM to generate tiny Dirac neutrino mass, neutrino magnetic moment can be expressed as  \cite{Fujikawa:1980yx}
\begin{equation}
    \mu_\nu = \frac{ e G_F m_\nu}{8 \sqrt{2}\pi^2} = 3 \times 10^{-20} \mu_B\, \left(\frac{m_\nu}{0.1~{\rm eV}}\right)~,
    \label{dirac}
\end{equation} 
     and it can be as large as $\sim 10^{-20}~\mu_B$. In the standard seesaw scenario, where neutrinos are Majorana particles,  the transition magnetic moments as a result of Standard Model interactions are given by
      \cite{Pal:1981rm}
\begin{equation}
    \mu_{ij} = -\frac{3eG_F}{32\sqrt{2}\pi^2}(m_i \pm m_j)\sum_{\ell = e, \mu,\tau}U_{\ell i}^* U_{\ell j} \frac{m_\ell^2}{m_W^2}.
\end{equation}
   The strength of transition neutrino magnetic moment is even smaller and of the order of $\sim 10^{-23}~\mu_B$ in this scenario. In the minimal Left-Right symmetric model, neutrino magnetic moment can be comparatively enhanced due to non-standard interactions via $W_R^\pm$ gauge boson \cite{Giunti:2014ixa}. However, the current experimental limit on the mixing angle between $W_R^\pm$ and $W^\pm$ does not permit the strength to be greater than $\sim 10^{-14}~\mu_B$. Similar strength of neutrino magnetic moment can be achievable in $R$-parity-violating supersymmetric extensions \cite{Kim:1976gk}. It was recently systematically analyzed and demonstrated in Ref.~\cite{Babu:2020ivd} that if neutrino magnetic moments are measured at the current experimental sensitivity  $\mathcal{O}(10^{-11})~\mu_B$ level, neutrinos are most likely Majorana particles, and the models based on $SU(2)_H$ symmetry\cite{Babu:2020ivd, Babu:1989wn, Babu:1990wv}  fit well within this category. Neutrino mass models \cite{Zee:1980ai, Babu:1992vq, Babu:2019mfe} based on spin-symmetry mechanism \cite{Barr:1990um} can accommodate neutrino transition magnetic moments as large as  $\mathcal{O}(10^{-12})~\mu_B$ \cite{Babu:2020ivd}. All these predictions on neutrino magnetic moments in different neutrino mass models \cite{Babu:2020ivd, Lindner:2017uvt, Giunti:2014ixa} are shown in green coloured lines in Fig.~\ref{sum}. It is quite interesting to see that the investigation of neutrino magnetic moments can be an excellent tool in searching for the theory underlying the neutrino mass generating mechanism.
\end{itemize}

\begin{figure}[!t] 
  \centering
   \includegraphics[width=0.65\textwidth]{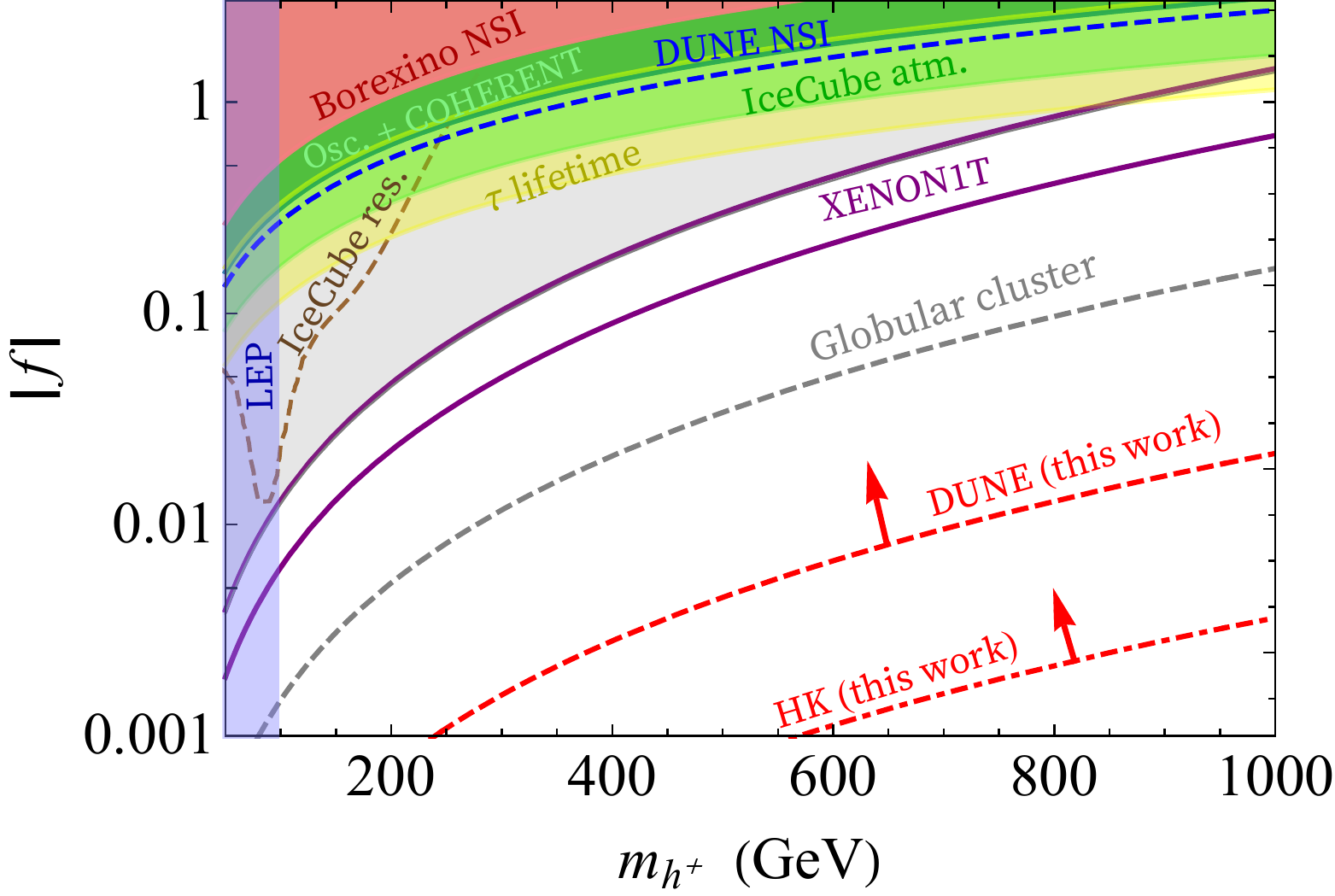}
  \caption{Different theoretical and experimental limits and future sensitivities on $SU(2)_H$ horizontal symmetric  model. See text for details. 
  }
  \label{pheno:su2h}
\end{figure}

Any UV complete model of large neutrino magnetic moments is expected to provide a more extensive phenomenology. For completeness of our study, we consider the model based on $SU(2)_H$ horizontal symmetry~\cite{Babu:2020ivd} which generates naturally large neutrino transition magnetic moments consistent with light neutrino masses and other existing experimental constraints. Several phenomenological consequences of this model have been recently studied in a different context~\cite{Babu:2020ivd, Babu:2021jnu}.
Following similar conventions and the same Lagrangian structure from Ref.~\cite{Babu:2020ivd}, we summarize the different phenomenological implications in the $SU(2)_H$  model parameter space in Fig.~\ref{pheno:su2h}. The transition magnetic moments of neutrino arise through charged Higgs induced quantum loop corrections. Due to SU(2)$_H$ symmetric limit, the contributions from two loop diagrams add up for neutrino magnetic moments, whereas it subtracts for neutrino mass contributions. In Fig.~\ref{pheno:su2h}, $h^+$ denotes the responsible charged scalar field, and $f$ is the relevant Yukawa term contributing to the neutrino magnetic moments. The colour shaded regions are excluded from current experimental searches: the light blue shaded region is excluded from charged Higgs searches at LEP experiment; the red shaded region is excluded from   Borexino NSI searches  \cite{Agarwalla:2019smc}; NSI from the global fit to neutrino oscillation data \cite{Esteban:2018ppq} constrains dark green shaded region, IceCube atmospheric data \cite{Esmaili:2013fva} on NSI imposes bound as indicated by light green shaded region; the yellow shaded region is bounded from $\tau $ lifetime constraints \cite{Babu:2020ivd}; the gray shaded region is excluded from neutrino magnetic moment searches at Borexino experiment \cite{Borexino:2017fbd}. There is an LEP mono-photon search limit \cite{Berezhiani:2001rs} as well; however, we find that it is comparatively less constrained than the above-mentioned scenarios.   We also project the future DUNE sensitivity \cite{Chatterjee:2021wac} from NSI searches by the blue dashed line. The brown dashed line represents the future sensitivity from IceCube by looking at Glashow-like resonance features \cite{Babu:2019vff, Babu:2022fje} induced by charged scalars in the ultrahigh-energy neutrino event spectrum. The gray dashed line is the astrophysical constraint originating from red giant and horizontal branch stars  $|\mu_\nu| \leq 1.5 \times 10^{-12} \mu_B$  (95\% CL)  \cite{Viaux:2013lha, Viaux:2013hca, Capozzi:2020cbu}. However, it has been recently pointed out that this astrophysical limit can be relaxed by considering ``neutrino trapping mechanism" \cite{Babu:2020ivd, Babu:2021jnu}. The red dashed lines indicate the future sensitivities on neutrino magnetic moments at DUNE and HK experiments based on our analysis and setting magnetic field strength $B_0 =10^{10}\,$G. We can see that a large region of the $SU(2)_H$ model parameter space can be explored by looking at imprints of neutrino magnetic moments on supernova neutrino signal at DUNE and HK.

\section{Conclusions }\label{SEC-07}
The neutronization burst from a future galactic SN has the potential to provide a wealth of information on neutrino transition magnetic moments (TMM). The neutrino spectra during this phase is usually dominated by a large fraction of $\nu_e$ with sub-dominant $\bar{\nu}_e$ and $\nu_{\mu,\tau}$. However, the presence of a non-zero neutrino TMM can change this picture completely. The combination of spin-flavour conversions due to a TMM, and resonant flavour conversions due to mass-mixings, can lead to a suppression of the $\nu_e$, while simultaneously enhancing the $\bar{\nu}_e$ spectra in this epoch. This tell-tale signature can be used to put strong bounds on neutrino TMMs using the spectra from a future galactic SN. With immense experimental effort underway in the detection of neutrinos from a future galactic SN, it is timely to analyze the impact of neutrino TMM on the signal.

In this work, we have studied the neutrino flavour evolution inside a SN in the presence of a finite TMM, and considered the effect it has on the spectra from the burst phase. By simulating the event spectra in upcoming neutrino experiments like as DUNE and Hyper-Kamiokande, we have found that the neutrino TMMs that may be probed by these experiments for a SN happening at 10$\,$kpc are two or three orders of magnitude (from $\mathcal{O} [10^{-11}]~ \mu_B$ to $\mathcal{O} [10^{-14}]~ \mu_B$ for the allowed range of magnetic field strengths) better than the current terrestrial and astrophysical bounds. We have discussed the uncertainties present in such an analysis, and the kind of impact it can have on our results. Furthermore, we have analyzed how this realization can shed light on three essential neutrino properties: (a) the Dirac/Majorana character of the neutrino, (b) mass ordering, and (c) the neutrino mass generation mechanisms. Finally, for completeness of our study, we have considered a model based on a $SU(2)_H$ horizontal symmetry, which generates large neutrino TMMs, and extrapolated our bounds onto the model parameter space. We have found that in such models, a large section of the parameter space can be explored using the neutrino signal from a future galactic SN.
\section*{Acknowledgments}

We thank Evgeny Akhmedov, Manfred Lindner, Orlando Peres and Suprabh Prakash for useful discussions. MS would like to thank Ivan Martinez-Soler for help with MARLEY simulations. YPPS thanks the   Max-Planck-Institut f{\"u}r Kernphysik Particle and Astroparticle Physics division for warm hospitality during the completion of this work.  The work of YPPS is supported in part by the FAPESP funding Grants No. 2014/19164-453 6, No. 2017/05515-0 and No. 2019/22961-9.

{\footnotesize
\bibliographystyle{utphys}
\bibliography{reference}}
\end{document}